\documentclass[11pt]{article}

\usepackage{geometry,setspace}
\usepackage{caption}
\usepackage{enumerate}
\usepackage{graphicx}
\usepackage{rotating,multirow}
\usepackage[table]{xcolor}
 \usepackage{amsmath,amsfonts,amssymb,mathrsfs,amsthm}
\usepackage{bbm}
\usepackage{bm}
\usepackage{natbib}
\usepackage{booktabs}
\usepackage{pdflscape}
\usepackage{todonotes}
\usepackage{authblk}
\usepackage{xr}
\usepackage{tikz-cd}
\usepackage{framed}

\usepackage{xfrac}

\usepackage{caption}
\usepackage[T1]{fontenc}
\input{glyphtounicode}
\newtheorem{proposition}{Proposition}
\newtheorem{theorem}{Theorem}
\newtheorem{lemma}{Lemma}

\theoremstyle{definition}
\newtheorem{definition}{Definition}
\newtheorem{remark}{Remark}
\newtheorem{algorithm}{Algorithm}

\newcommand{\indicator}[1]{\ensuremath{I_{\{#1\}}}}

\usepackage{accents}
\newcommand{\vtrans}{\ensuremath{\mathcal{V}}}

\newcommand{\downprob}{\ensuremath{\delta}}
\newcommand{\udual}{\ensuremath{u^*}}

\newcommand{\Downprob}{\ensuremath{\Delta}}

\renewcommand{\P}{\mathbb{P}}
\newcommand{\E}{\mathbb{E}}
\newcommand{\R}{\mathbb{R}}

\newcommand{\N}{\mathbb{N}}

\newcommand{\var}{\operatorname{var}}

\renewcommand{\leq}{\leqslant}
\renewcommand{\geq}{\geqslant}

\newcommand{\rd}{\mathrm{d}}

\newcommand{\myThanks}{\thanks{Address correspondence to
     Alexander J.~McNeil, The York Management School,
     University of York, Freboys Lane, York YO10 5GD, UK, +44 (0) 1904 325307,
     \texttt{alexander.mcneil@york.ac.uk}.}}

\begin{document}

\title{Modelling volatile time series with v-transforms and copulas}
\author{Alexander J. McNeil\myThanks}
\affil{The York Management School, University of York}

\date{12th January 2021}

\maketitle
\begin{abstract}
An approach to the modelling of volatile time series using a class
of uniformity-preserving transforms for uniform random variables is proposed. V-transforms
describe the relationship between quantiles of the stationary
distribution of the time series
and quantiles of the distribution of a predictable volatility proxy
variable. They can be represented as copulas and permit the
formulation and estimation of models that combine arbitrary marginal
distributions with copula processes for the
dynamics of the volatility proxy. The idea is illustrated using a
Gaussian ARMA copula process and the resulting model is shown to
replicate many of the stylized facts of financial return series and to
facilitate the calculation of marginal and conditional characteristics
of the model including quantile measures of risk.
Estimation is carried out by adapting the exact maximum
likelihood approach to the estimation of ARMA processes and the
model is shown to be competitive with standard GARCH in an
empirical application to Bitcoin return data.
\end{abstract}


\section{Introduction}\label{sec:intro}

In this paper, we show that a class of uniformity-preserving transformations
for uniform random variables can facilitate the
application of copula modelling to time series exhibiting the serial
dependence characteristics that are typical of volatile financial
return data. Our main aims are twofold: to
establish the fundamental properties
of v-transforms and show that they are a natural fit to the
volatility modelling problem; to develop a class of processes using the implied
copula process of a Gaussian ARMA model that can serve as an archetype for 
copula models using v-transforms.
Although the existing literature on volatility modelling in econometrics
is vast, the models we propose have some attractive features. In
particular, as
copula-based models, they allow the separation of marginal and
serial dependence behaviour in the construction and estimation of models.


A distinction is commonly made between genuine
stochastic volatility models, as investigated by~\cite{bib:taylor-94}
and~\cite{bib:andersen-94}, and GARCH-type models as developed in a
long series of papers
by~\cite{bib:engle-82},~\cite{bib:bollerslev-86},~\cite{bib:ding-engle-granger-93},~\cite{bib:glosten-jagannathan-runkle-93}
and~\cite{bib:bollerslev-engle-nelson-94}, among others. In the former
an unobservable process describes the
volatility at any time point while in the latter
volatility is modelled as a function of observable information describing the past
behaviour of the process; see also the review articles
by~\citet{bib:shephard-96} and~\citet{bib:andersen-benzoni-10}. The
generalized autoregressive score (GAS) models
of~\citet{bib:creal-koopman-lucas-13} generalize the observation-driven
approach of GARCH models by using the score function of the
conditional density to model time variation in key parameters of the
time series model. The models of this paper have more in common with the observation-driven
approach of GARCH and GAS but have some important differences.

In GARCH-type models, the marginal distribution of a stationary process
is inextricably linked to the dynamics of the process as well as the
conditional or innovation
distribution; in most cases, it has no
simple closed form. For example, the standard GARCH mechanism serves to create power-law behaviour in
the marginal distribution, even when the innovations come from
a lighter-tailed distribution such as
Gaussian~\citep{bib:mikosch-starica-00}. While such models work
well for many return series, they may not be sufficiently flexible to
describe all possible combinations of marginal and serial dependence
behaviour encountered in applications. In the empirical example of this paper, which relates to
log-returns on the Bitcoin price series, the data appear to
favour a marginal distribution with sub-exponential tails that are
lighter than power tails and this cannot be well captured by standard
GARCH models. Moreover, in contrast to much
of the GARCH literature, the models we propose make no
assumptions about the existence of second-order moments and could also
be applied to very heavy-tailed situations where variance-based
methods fail.


Let $X_1,\ldots,X_n$ be a time series of financial returns sampled at (say)
daily frequency and assume that these are modelled by a strictly stationary stochastic process $(X_t)$ with marginal distribution function (cdf) $F_X$. 
To match the
stylized facts of financial return data described, for example,
by~\cite{bib:campbell-lo-mackinlay-97} and~\cite{bib:cont-01}, it is generally
agreed that 
$(X_t)$ should have limited serial correlation, but the
squared or absolute processes $(X_t^2)$ and $(|X_t|)$ should have
significant and persistent positive serial correlation to describe the
effects of volatility clustering. 

In this paper, we refer to transformed series like $(|X_t|)$, in which
volatility is revealed through serial correlation, as
\textit{volatility proxy series}. More generally, a volatility proxy
series $(T(X_t))$ is obtained by applying a transformation $T:\R
\mapsto \R$ which (i) depends on a change point $\mu_T$ that may be
zero, (ii) is increasing in $X_t-\mu_T$ for $X_t \geq \mu_T$ and (iii)
is increasing in $\mu_T - X_t$ for $X_t \leq \mu_T$.

Our approach in this paper is to model the probability-integral
transform (PIT) series $(V_t)$ of a volatility proxy series. This is defined by
$V_t = F_{T(X)}(T(X_t))$ for all $t$, where
$F_{T(X)}$ denotes the cdf of $T(X_t)$. If $(U_t)$ is the PIT series of the original process $(X_t)$, defined by $U_t = F_X(X_t)$ for all $t$, then a 
\textit{v-transform} is a function describing the relationship between the terms of 
$(V_t)$ and the terms of
 $(U_t)$.  
 Equivalently, a v-transform describes the relationship between quantiles of the distribution of $X_t$ and the distribution of the volatility proxy $T(X_t)$. Alternatively, it characterizes the dependence structure or copula of the pair of variables $(X_t, T(X_t))$.
 In this paper, we show how to derive flexible,
parametric families of v-transforms for practical modelling purposes. 
 
\begin{figure}[htb]
  \centering
   \includegraphics[width=17cm,height=7cm]{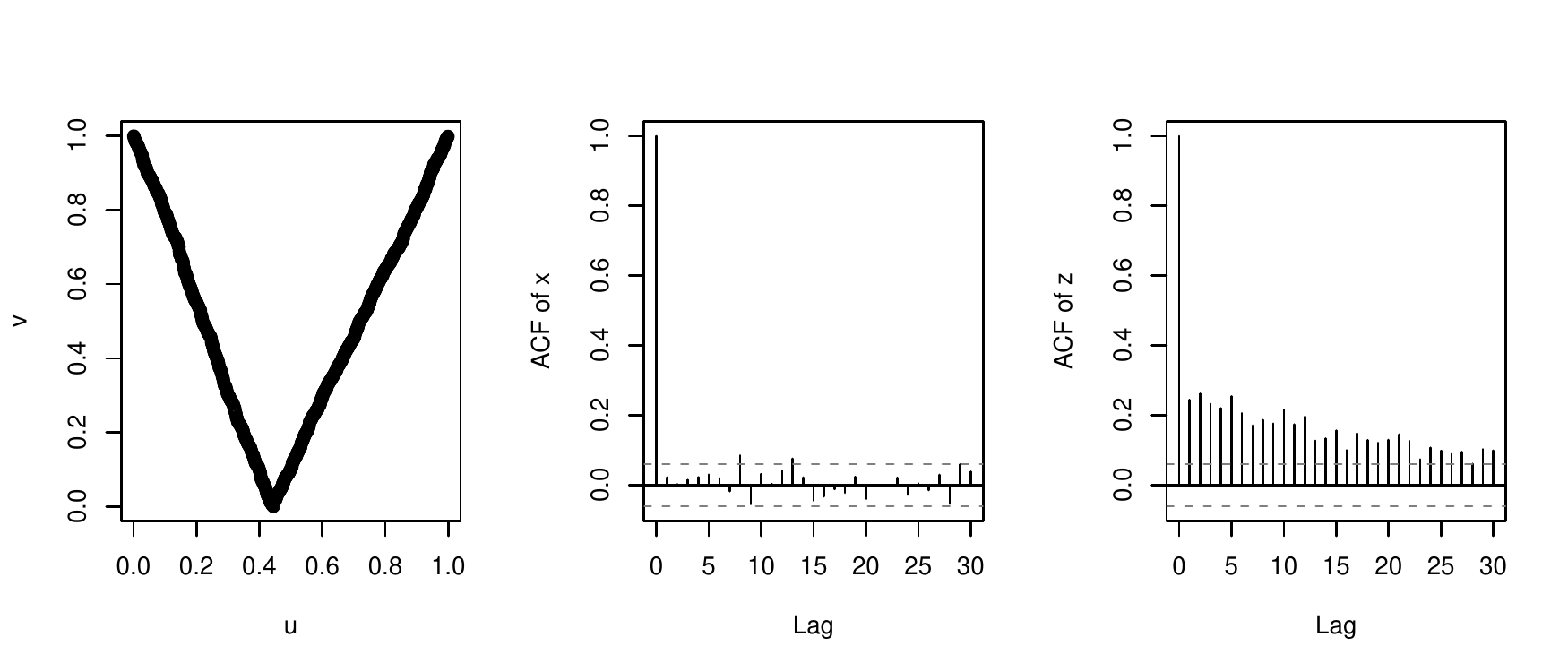}
   \caption{\label{fig:1}  {Scatterplot of
$v_t$ against $u_t$ (left), sample acf of raw data $x_t$ (centre) and sample
acf of $z_t =\Phi^{-1}(v_t)$ (right). The transformed data are defined by $v_t =F^{(|X|)}_n(|x_t|) $ and $u_t =
F^{(X)}_n(x_t)$ where $F_n^{(X)}$ and $F_n^{(|X|)}$
denote versions of the empirical distribution function of the $x_t$ and $|x_t|$
values, respectively. The sample size is $n=1043$ and the data are
daily log-returns of the Bitcoin price for the years
2016--2019. }}  
 \end{figure}

 To gain insight into the typical form of a v-transform, let
 $x_1,\ldots,x_n$ represent the realized data values and let $u_1,\ldots, u_n$ and $v_1,\ldots, v_n$ be the samples obtained by
  applying the transformations $v_t =F^{(|X|)}_n(|x_t|) $ and $u_t =
F^{(X)}_n(x_t)$, where $F_n^{(X)}(x)=\frac{1}{n+1}\sum_{t=1}^n \indicator{x_t \leq x}$ and $F_n^{(|X|)}(x)=\frac{1}{n+1}\sum_{t=1}^n \indicator{|x_t| \leq x}$
denote scaled versions of the empirical distribution functions of the $x_t$ and $|x_t|$
samples, respectively. The graph of $(u_t, v_t)$ gives an empirical
estimate of the v-transform for the random variables $(X_t,|X_t|)$.
In the left-hand plot of Figure~\ref{fig:1} we show the relationship
for a sample of $n=1043$ daily log-returns of the Bitcoin price series
for
the years 2016--2019. Note how the empirical v-transform takes
the form of a slightly asymmetric `V'.

The right-hand plot of Figure~\ref{fig:1} shows the sample
autocorrelation function (acf) of the data given by $z_t =
\Phi^{-1}(v_t)$ where $\Phi$ is the standard normal cdf.  This reveals 
a persistent pattern of positive serial correlation which can be
modelled by the implied ARMA copula.  {This pattern is not
  evident in the acf of the raw $x_t$ data in the centre plot.}

  To construct a volatility model for $(X_t)$ using v-transforms, we need to specify a process for $(V_t)$.
In principle, any model for a series of serially dependent uniform
variables can be applied to $(V_t)$. In this paper, we illustrate concepts
using the Gaussian copula model implied by the standard ARMA
dependence structure. This model is particularly tractable and allows
us to derive model properties and fit models to data relatively easily.

There is a large literature on copula models for time series; see, for example, the review papers
by~\citet{bib:patton-12} and~\citet{bib:fan-patton-14}. While the
main focus of this literature has been on cross-sectional dependencies
between series, there is a growing literature on models of serial
dependence. First-order Markov copula models have been investigated
by~\citet{bib:chen-fan-06b},~\cite{bib:chen-wu-yi-09}
and~\cite{bib:domma-giordano-perri-09} while higher-order Markov
copula models using D-vines are applied
by~\citet{bib:smith-min-almeida-czado-10}. These models are
based on the
pair-copula apporoach developed in~\cite{bib:joe-96}, 
~\citeauthor{bib:bedford-cooke-01b} (\citeyear{bib:bedford-cooke-01b},
\citeyear{bib:bedford-cooke-02}) and
\cite{bib:aas-czado-frigessi-bakken-09}. 
However, the standard bivariate 
copulas that enter these models are not generally effective at
describing the typical
serial dependencies created by stochastic volatility, as 
observed by~\cite{bib:louaiza-maya-et-al-18}.

The paper is structured as follows. In
Section~\ref{sec:simple-symm-model}, we provide motivation for the paper by constructing a symmetric model using the simplest example of a v-transform. The general theory of v-transforms is developed in Section~\ref{sec:v-transforms} and is used to construct the class of VT-ARMA processes and analyse their properties in Section~\ref{sec:properties-model}. Section~\ref{sec:estimation-model}
treats estimation and statistical inference for VT-ARMA processes and
provides an example of their application to the Bitcoin return data;
Section~\ref{sec:conclusion} presents the conclusions. Proofs may be found in the Appendix \ref{sec:proofs}.

\section{A motivating model}\label{sec:simple-symm-model}

Given a probability space $(\Omega,\mathcal{F},\P)$, we construct a symmetric, strictly stationary
process $(X_t)_{t\in\N\setminus\{0\}}$ such that, under the even
transformation $T(x)=|x|$, the serial
dependence in the volatility proxy series $(T(X_t))$ is of ARMA type. We assume that the marginal cdf $F_X$ of $(X_t)$ is absolutely continuous and the density $f_X$ satisfies $f_X(x) = f_X(-x)$ for all $x >0$. Since $F_X$ and $F_{|X|}$ are both continuous the properties of the probability-integral (PIT) transform imply that the series $(U_t)$ and $(V_t)$ given by $U_t = F_X(X_t)$ and $V_t = F_{|X|} (|X_t|)$ both have standard
uniform marginal distributions. Henceforth we refer to $(V_t)$ as the
\textit{volatility PIT process} and $(U_t)$ as the \textit{series PIT process}.

Any other volatility proxy
series that can be obtained by a continuous and strictly increasing transformation
of the terms of $(|X_t|)$, such as $(X_t^2)$, yields exactly the same volatility PIT process.
For example, if $\tilde{V}_t=
F_{X^2}(X_t^2)$, then it follows from the fact that $F_{X^2}(x) =
F_{|X|}(\sqrt[+]{x})$ for $x\geq 0$ that
 $\tilde{V}_t=F_{X^2}(X_t^2) = F_{|X|} (|X_t|)=V_t$. In this sense we
 can think of classes of equivalent volatility proxies, such as
 $(|X_t|)$, $(X_t^2)$, $(\exp|X_t|)$ and $(\ln(1+|X_t|))$. In fact
 $(V_t)$ is itself an equivalent volatility proxy  to $(|X_t|)$ since
 $F_{|X|}$ is a continuous and strictly increasing transformation.

The symmetry of $f_X$ implies that $F_{|X|}(x) = 2F_X(x)-1 = 1-2F_X(-x)$ for $x\geq 0$. Hence
we find that
\begin{equation*}
V_t = F_{|X|} (|X_t|) =
\begin{cases}
\begin{aligned}
F_{|X|}(-X_t) &= 1-2F_X(X_t) =1 -2U_t, &\text{if $X_t <0$}\\
F_{|X|}(X_t) &= 2F_X(X_t)-1 = 2U_t -1, &\text{if $X_t \geq 0$}
\end{aligned}
\end{cases}
\end{equation*}
which implies 
that the relationship between the volatility PIT process
$(V_t)$ and the series PIT process $(U_t)$ is given by
\begin{equation}\label{eq:13}
  V_t = \vtrans(U_t) =|2 U_t - 1| 
\end{equation}
where $\vtrans(u) =
|2u-1|$ is a perfectly symmetric v-shaped function that maps values of $U_t$ close to 0
or 1 
to values of $V_t$
close to 1, and values close to $0.5$ to values close to
0. $\mathcal{V}$ is the canonical example of a v-transform. It is
related to the so-called tent-map transformation $\mathcal{T}(u)=2\min(u,1-u)$ by $\mathcal{V}(u)=1-\mathcal{T}(u)$.

Given $(V_t)$, let the process
$(Z_t)$ be defined by setting $Z_t =
\Phi^{-1}(V_t)$ so that we have the following chain of transformations
\begin{equation}\label{eq:23}
\begin{tikzcd}
 X_t \ar{r}{F_X} & U_t \ar{r}{\vtrans}  & V_t
 \ar{r}{\Phi^{-1}}  & Z_t\quad .
\end{tikzcd}
\end{equation}
We refer to $(Z_t)$ as a \textit{normalized volatility proxy series}. Our aim is to construct a process $(X_t)$
such that, under the chain of 
transformations in~\eqref{eq:23}, we obtain a Gaussian ARMA process $(Z_t)$ with
mean zero and variance one. We do this by working back through the
chain.

The transformation $\vtrans$ is not an
injection and, for any $V_t>0$, there are two possible inverse values, $\tfrac{1}{2}(1-V_t)$ and $\tfrac{1}{2}(1+V_t)$. However, by randomly choosing between these values, we can `stochastically invert' 
$\vtrans$ to construct a random variable $U_t$ such that $\vtrans(U_t)
= V_t$, This is summarized in Lemma~\ref{lemma:reconstructU}, which is a special case of a more general result in
Proposition~\ref{prop:resconstructU}. 
\begin{lemma}\label{lemma:reconstructU}
Let $V$ be a standard uniform variable.  If $V=0$ set $U =
\tfrac{1}{2}$. Otherwise let $U = \tfrac{1}{2}(1-V)$ with probability 0.5
and $U = \tfrac{1}{2}(1+V)$ with probability 0.5.
Then $U$ is uniformly distributed and
$\vtrans(U) = V$.
\end{lemma}
This simple result suggests the following algorithm for constructing a 
process $(X_t)$ with symmetric marginal density $f_X$ such that the
corresponding normalized volatility proxy process $(Z_t)$ under the absolute
value transformation (or continuous and strictly increasing functions thereof) is an ARMA process. We describe the resulting model as a VT-ARMA process.
\begin{framed}
\begin{algorithm}\label{algo1}
\begin{enumerate}
\item Generate $(Z_t)$ as a causal and invertible Gaussian ARMA process of order $(p,q)$
  with mean zero and variance one.
\item Form the volatility PIT process $(V_t)$ where $V_t =\Phi(Z_t)$
  for all $t$.
\item Generate a process of iid Bernoulli variables $(Y_t)$ such that
  $\P(Y_t=1) = 0.5$.
\item Form the PIT process $(U_t)$ using the transformation
   $ U_t =0.5 (1-V_t)^{\indicator{Y_t = 0}} (1+V_t)^{\indicator{Y_t =
        1}}$.
\item Form the process $(X_t)$ by setting $X_t = F_X^{-1}(U_t)$.
\end{enumerate}
\end{algorithm}
\end{framed}
It is important to state that the use of the Gaussian process $(Z_t)$
as the fundamental building block of the VT-ARMA process in
Algorithm~\ref{algo1} has no effect on the marginal distribution of
$(X_t)$, which is $F_X$ as specified in the final step of the
algorithm. The process $(Z_t)$ is exploited \textit{only for its
  serial dependence structure}, which is described by a family of
finite-dimensional Gaussian copulas; this dependence structure is
applied to the volatility proxy process.

\begin{figure}[htb]
  \centering
   \includegraphics[width=14cm,height=10cm]{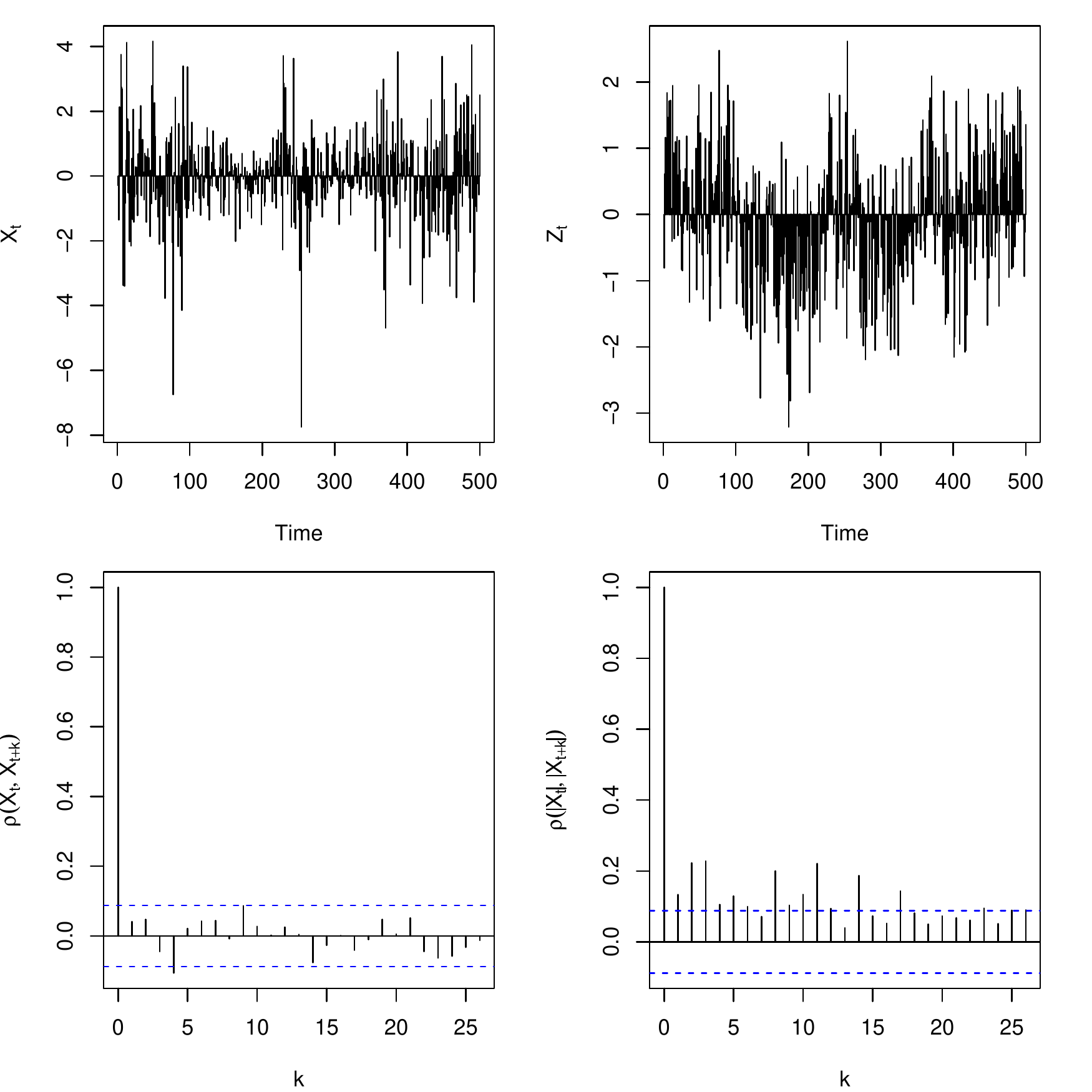}
   \caption{\label{fig:3A} Realizations of length $n=500$ of $(X_t)$ and $(Z_t)$ for a VT-ARMA(1,1) process with a marginal Student t distribution with $\nu=3$ degrees of freedom and ARMA paramaters $\alpha=0.95$ and $\beta =-0.85$. ACF plots for $(X_t)$ and $(|X_t|)$ are also shown.}
 \end{figure}

 Figure~\ref{fig:3A} shows a symmetric VT-ARMA(1,1) process with ARMA parameters
 $\alpha_1 = 0.95$ and $\beta_1 = -0.85$; such a model often works
 well for financial return data. Some intuition for this observation
 can be gained from the fact that the popular GARCH(1,1) model is
 known to have the structure of an ARMA(1,1)
 model for the squared data process; see, for example,~\citet{bib:mcneil-frey-embrechts-15} (Section 4.2) for more details.

\section{V-transforms}\label{sec:v-transforms}
To generalize the class of v-transforms we admit two forms of asymmetry in the construction described in Section~\ref{sec:simple-symm-model}: we allow the
density $f_X$ to be skewed; we introduce an asymmetric volatility
proxy.
\begin{definition}[Volatility proxy transformation and profile]
Let $T_1$ and $T_2$ be strictly
increasing, continuous and differentiable functions on $\R^+=[0,\infty)$ such that $T_1(0) = T_2(0)$. Let $\mu_T \in \R$. Any transformation $T:\R \to \R$ of the form
\begin{equation}
  \label{eq:9}
  T(x) = 
\begin{cases}
T_1(\mu_T-x)&\quad x \leq \mu_T\\
T_2(x-\mu_T) &\quad x > \mu_T
\end{cases}
\end{equation}
is a volatility proxy transformation. The parameter $\mu_T$ is the \textit{change point} of $T$ and the associated
function $g_T:\R^+\to\R^+$, $g_T(x) = T_2^{-1} \circ
T_1(x)$ is the \textit{profile function} of $T$.
\end{definition}
By introducing $\mu_T$ we allow the possibility that the natural change point may not be identical to zero. By introducing different functions $T_1$ and $T_2$ for returns on either side of the change point, we allow the possibility that one or other may contribute
more to the volatility proxy. This has a similar
economic motivation to the \textit{leverage} effects in GARCH models~\citep{bib:ding-engle-granger-93}; falls in
equity prices increase a firm's leverage and increase the volatility
of the share price. 

Clearly the profile function of a volatility proxy transformation is a strictly increasing,
continuous and differentiable function on $\R^+$ such that $g_T(x) =
0$. In conjunction with $\mu_T$, the profile contains all the
information about $T$ that is relevant for constructing
v-transforms. In the
case of a volatility proxy transformation that is symmetric about
$\mu_T$, the profile satisfies $g_T(x) = x$.

The following result shows how v-transforms $V = \vtrans(U)$ can be obtained by considering
different continuous distributions $F_X$ and different volatility proxy transformations
$T$ of type~(\ref{eq:9}).

\begin{proposition}\label{prop:model-with-skewness}
Let $X$ be a random variable with absolutely continuous and strictly increasing cdf $F_X$ on
$\R$ and let $T$ be a volatility proxy transformation. Let $U=
F_X(X)$ and $V= F_{T(X)}(T(X))$.  Then $V =\vtrans(U)$ where
  \begin{equation}
    \label{eq:4}
    \vtrans(u) =
\begin{cases}
F_X\left(\mu_T + g_T\left(\mu_T-F_X^{-1}(u)\right)\right) -
u, &u \leq F_X(\mu_T) \\
 u - F_X\left(\mu_T- g_T^{-1}\left( F_X^{-1}(u) -\mu_T \right)\right) ,& u > F_X(\mu_T)\,.
\end{cases}
  \end{equation}
\end{proposition}  

The result implies that any two volatility proxy transformations
$T$ and $\tilde{T}$ which have the same change point $\mu_T$ and
profile function $g_T$ belong to an equivalence class with respect to the
resulting v-transform. This generalizes the idea that $T(x) =|x|$ and
$T(x) =x^2$ give the same v-transform in the symmetric case of
Section~\ref{sec:simple-symm-model}.
Note also that the volatility proxy transformations $T^{(V)}$ and
$T^{(Z)}$ defined by
\begin{eqnarray}
  T^{(V)}(x)  & = &
                    F_{T(X)}(T(x)) = \vtrans\big( F_X(x) \big)
                    \nonumber \\
  T^{(Z)}(x) & =  & \Phi^{-1}(T^{(V)}(x)) = \Phi^{-1} \Big( \vtrans\big(
                    F_X(x) \big) \Big)\label{eq:volproxyTz}
\end{eqnarray}
are in the same equivalence class as $T$ since they share the same
change point and
profile function.

\begin{definition}[v-transform and fulcrum]\label{def:v-transforms-1}
Any transformation $\vtrans$ that can be obtained from equation~(\ref{eq:4})
by choosing an absolutely
continuous and strictly increasing cdf $F_X$ on $\R$ and a volatility proxy transformation $T$
is a v-transform. The value $\downprob = F_X(\mu_T)$ is the
\textit{fulcrum} of the v-transform.
\end{definition}

\subsection{A flexible parametric family}\label{sec:flex-param-family}

In this
section we derive a family of v-transforms using
construction~(\ref{eq:4}) by taking a tractable asymmetric model for
$F_X$ using the construction proposed by~\citet{bib:fernandez-steel-98} and
by setting $\mu_T=0$ and $g_T(x) = k x^\xi$ for $k>0$ and
$\xi>0$. This profile function contains the identity
profile $g_T(x) = x$ (corresponding to the symmetric volatility proxy transformation)
as a special case, but allows cases
where negative or positive returns contribute more to the
volatility proxy. The choices we make may at first sight seem rather
arbitrary, but the resulting family can in fact assume many of the 
shapes that are permissable for v-transforms, as we will argue.

Let $f_0$ be a density that is symmetric about the origin
and let $\gamma>0$ be a scalar parameter. Fernandez and Steel suggested the model
\begin{equation}
  \label{eq:10}
\  f_X(x ;\gamma) = 
\begin{cases}
\frac{2\gamma}{1+\gamma^2}\; f_0(\gamma x) &\quad x \leq 0 \\
\frac{2\gamma}{1+ \gamma^2}\; f_0\left(\frac{x}{\gamma}\right)&\quad x > 0\,.
\end{cases}
\end{equation}
This model is often used to obtain skewed normal and skewed Student
distributions for use as innovation distributions in econometric
models. 
A model with
$\gamma > 1$ is skewed to the right while a model with $\gamma < 1$
is skewed to the left, as might be expected for asset returns.
We consider the particular case of a Laplace or double exponential distribution $f_0(x) = 0.5
\exp(-|x|)$ which leads to particularly tractable expressions.
\begin{proposition}\label{prop:parametric-family}
Let $F_X(x;\gamma)$ be the cdf of the density~\eqref{eq:10} when
$f_0(x) = 0.5
\exp(-|x|)$. Set $\mu_T=0$ and let $g_T(x) = k x^\xi$ for
$k,\xi>0$. The v-transform~(\ref{eq:4}) is
given by
\begin{equation}
  \label{eq:11}
 \vtrans_{\downprob,\kappa,\xi}(u) = 
\begin{cases}
  1-u -
  (1-\downprob)\exp\left(-\kappa \left( -\ln\left(\frac{u}{\downprob}\right) \right)^\xi
  \right)&\quad u
 \leq \downprob,\\
u -
\downprob \exp\left( - \kappa^{-1/\xi} \left( -\ln\left(\frac{1-u}{1-\downprob}   \right)   \right)^{1/\xi}\right)& \quad u > \downprob,
\end{cases}
\end{equation}
where $\downprob = F_X(0) =(1+\gamma^2)^{-1} \in (0,1)$ and $\kappa =
k/\gamma^{\xi+1} > 0$. 
\end{proposition}

It is remarkable that~\eqref{eq:11} is a uniformity-preserving
transformation. If we set $\xi=1$ and $\kappa=1$ we get
\begin{equation}
  \label{eq:1}
 \vtrans_\downprob(u) =
\begin{cases}
 (\downprob-u)/\downprob& \quad u \leq \downprob,\\
 (u-\downprob)/(1-\downprob)& \quad u > \downprob
\end{cases}
\end{equation}
which obviously includes the symmetric
model $\vtrans_{0.5}(u) = |2u -1|$.
The v-transform $\vtrans_\downprob(u)$ in~\eqref{eq:1} is a
very convenient special case and we refer to it as the \textit{linear}
v-transform.

\begin{figure}[htb]
  \centering
   \includegraphics[width=12cm,height=12cm]{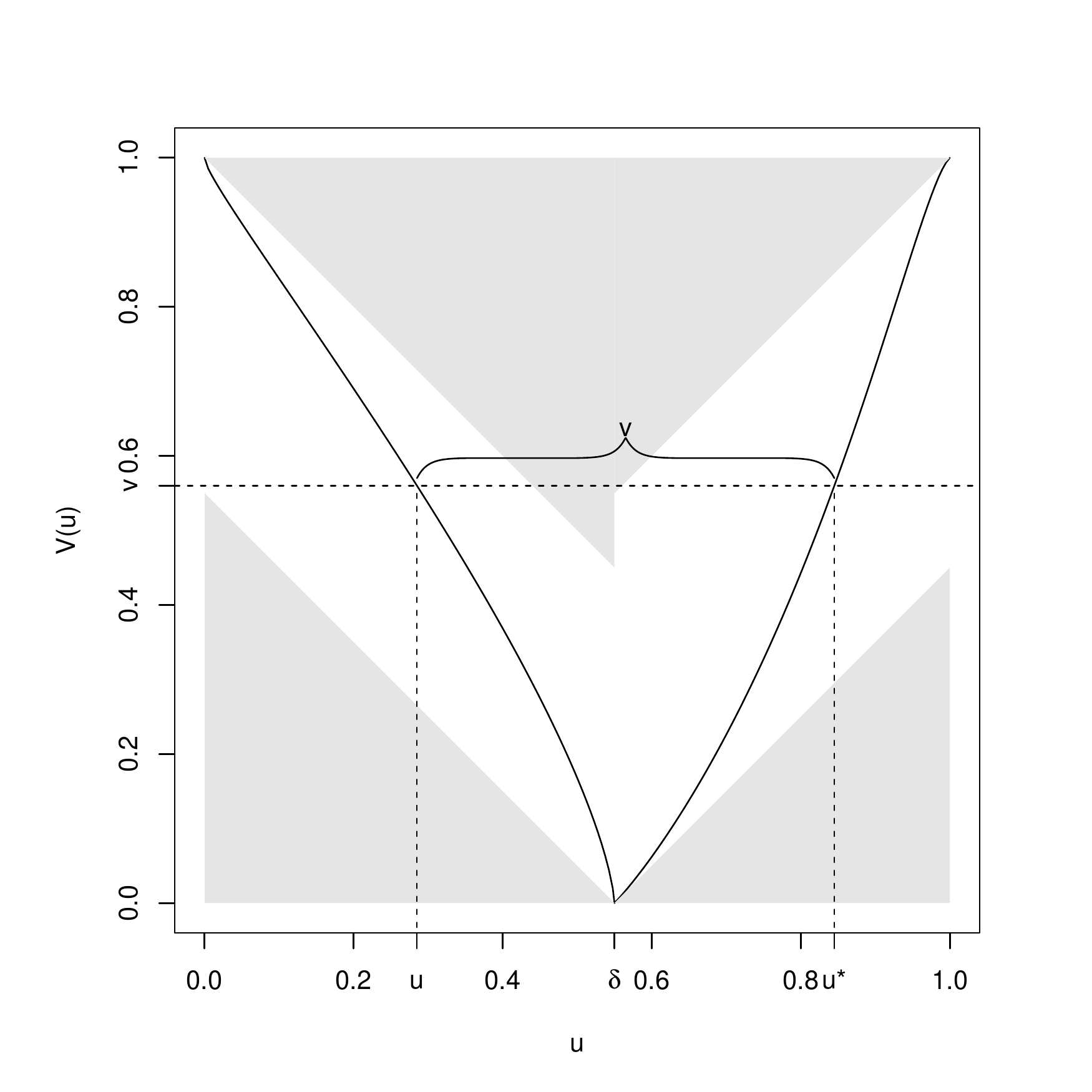}
   \caption{\label{fig:99} An asymmetric v-transform from the family
  defined in~(\ref{eq:11}). For any v-transform, if $v =\vtrans(u)$ and $u^*$ is the dual of $u$, then the points $(u,0)$, $(u,v)$, $(\udual,0)$ and $(\udual,v)$ form the vertices of a square. For the given fulcrum $\downprob$, a v-transform can never enter the gray shaded area of the plot. }
 \end{figure}

In Figure~\ref{fig:99} we show the v-transform
$\vtrans_{\downprob,\kappa,\xi}$ when $\downprob = 0.55$, $\kappa =
1.4$ and $\xi = 0.65$. We will use this particular v-transform to illustrate further properties of v-transforms and
find a characterization.

 \subsection{Characterizing v-transforms}

 It is easily verified that any v-transform obtained from~(\ref{eq:4}) consists of two arms or
 branches, described by continuous and strictly monotonic functions;
 the left arm is decreasing and the right arm increasing. See
 Figure~\ref{fig:99} for an illustration.
At the fulcrum $\downprob$ we have $\vtrans(\downprob) = 0$. Every
point $u \in [0,1]\setminus \{\downprob\}$ has a \textit{dual point}
$\udual$ on the opposite side of the fulcrum such that
$\vtrans(\udual) =\vtrans(u)$. Dual points can be interpreted as the
quantile probability levels of the distribution of $X$ that give rise to the same level of volatility.

We collect these properties together in the following lemma and add
one further important property that we refer to as  the \textit{square property}
of a v-transform; this property places constraints on the shape
that v-transforms can take and is illustrated in Figure~\ref{fig:99}.

\begin{lemma}\label{cor:vtranstions}
A v-transform is a mapping $\vtrans:[0,1] \to [0,1]$ with the
following properties:
  \begin{enumerate}
\item $\vtrans(0) = \vtrans(1) = 1$;
 \item There exists a point $\downprob$ known as the fulcrum
   such that $0 < \downprob < 1$ and $\vtrans(\downprob) = 0$;
 \item $\vtrans$ is continuous;
\item $\vtrans$ is strictly decreasing on
     $[0,\downprob]$ and strictly increasing on
       $[\downprob, 1]$;
       \item Every point $u \in [0,1]\setminus \{\downprob\}$ has a
         dual point $u^*$ on the opposite
         side of the fulcrum satisfying $\vtrans(u) = \vtrans(u^*)$
         and 
         $|u^* - u | = \vtrans(u)$ (square property).
\end{enumerate}
\end{lemma}
\noindent 
It is instructive to see why the square property must hold. Consider
Figure~\ref{fig:99} and fix a point $u \in [0,1]
\setminus \{\downprob\}$ with $\vtrans(u) = v$. Let $U\sim U(0,1)$ and
let $V  =\vtrans(U)$. The events $\left\{V \leq v\right\}$
 and $\left\{ \min(u,\udual) \leq U \leq \max(u,\udual)\right\}
 $ are the same and hence the uniformity of $V$ under a v-transform implies that
\begin{equation}
  \label{eq:3}
  v = \P(V \leq v) = \P \left(\min(u,\udual) \leq U \leq
    \max(u,\udual) \right) =
  |\udual -u| \,.
\end{equation}

The properties in Lemma~\ref{cor:vtranstions} could be taken
as the basis of an alternative definition of a v-transform. In view
of~\eqref{eq:3} it is clear that any mapping
$\vtrans$ that has these properties is a uniformity-preserving
transformation. We can characterize the mappings $\vtrans$ that have
these properties as follows.

\begin{theorem}\label{theorem:v-characterization}
A mapping $\vtrans: [0,1] \to [0,1]$ has the properties listed in Lemma~\ref{cor:vtranstions}  if and only if it takes the form
  \begin{equation}
    \label{eq:2}
    \vtrans(u) =
    \begin{cases}
(1-u) - (1-\downprob) \Psi \left( \frac{u}{\downprob} \right) & u \leq
\downprob, \\
u - \downprob \Psi^{-1}\left( \frac{1-u}{1-\downprob} \right) & u > \downprob,
\end{cases}
\end{equation}
where $\Psi$ is a continuous and strictly increasing distribution
function on $[0,1]$.
\end{theorem}

Our arguments so far show that every v-transform must have the form~\eqref{eq:2}. It remains to
verify that every uniformity-preserving transformation of the
form~\eqref{eq:2} can be obtained from construction~(\ref{eq:4}) and
this is the purpose of the final result of this section. This
allows us to view Definition~\ref{def:v-transforms-1}, Lemma~\ref{cor:vtranstions} and the
characterization~\eqref{eq:2} as three equivalent approaches to the
definition of v-transforms. 

\begin{proposition}\label{prop:reconstructT}
Let $\vtrans$ be a uniformity-preserving transformation  of the
form~\eqref{eq:2} and $F_X$ a continuous distribution function. Then
$\vtrans$ can be obtained from construction~\eqref{eq:4} using any volatility proxy transformation with change point $\mu_T = F_X^{-1}(\delta)$ and profile 
\begin{equation}\label{eq:volprofile}
g_T(x)  =  F_X^{-1}\left( F_X(\mu_T-x) + \vtrans\left(  F_X(\mu_T-x) \right)\right) -\mu_T, \quad x \geq 0.
\end{equation}
\end{proposition}

Henceforth we can view~\eqref{eq:2} as the general equation of a
v-transform. Distribution functions $\Psi$ on $[0,1]$ can be thought
of as \textit{generators} of v-transforms.
Comparing~\eqref{eq:2} with~(\ref{eq:11}) we see that our parametric
family $\vtrans_{\downprob,\kappa,\xi}$
is generated by $\Psi(x) = \exp(-\kappa(-
(\ln x)^\xi))$. This is a 2-parameter
distribution whose density can assume many different shapes on the unit
interval including increasing, decreasing, unimodal and bathtub-shaped
forms. In this respect it is quite similar to the beta distribution
which would yield an alternative family of v-transforms. The uniform
distribution function $\Psi(x) = x$ gives the family of linear
v-transforms $\vtrans_\downprob$.

In applications we construct models starting
from the building blocks of a tractable v-transform $\vtrans$ such
as~\eqref{eq:11} and a distribution
$F_X$; from these we can always infer an implied profile function
$g_T$ using~\eqref{eq:volprofile}. The alternative approach of starting from $g_T$ and  $F_X$ and
constructing $\vtrans$ via~\eqref{eq:4} is also possible but can lead to v-transforms that
are cumbersome and computationally expensive to evaluate if $F_X$ and
its inverse do not have simple closed forms.

\subsection{V-transforms and copulas}

If two uniform random variables are linked by the v-transform $V =
\vtrans(U)$ then the joint distribution function of $(U,V)$ is a
special kind of copula. In this
section we derive the form of the copula, which facilitates the
construction of stochastic processes using v-transforms.

To state the main result we use the notation $\vtrans^{-1}$ and
$\vtrans^\prime$ for the the inverse function and the gradient function of a v-transform
$\vtrans$. Although there is no unique inverse $\vtrans^{-1}(v)$
(except when $v=0$) the fact that the two branches of a v-transform
mutually determine each other allows us to define $\vtrans^{-1}(v)$ to
be the inverse of the
left branch of the v-transform given by $\vtrans^{-1}: [0,1] \to [0,\delta], \;
\vtrans^{-1}(v) = \inf\{u :\vtrans(u) = v\}$. The gradient
$\vtrans^\prime(u)$ is defined
for all points $u \in [0,1] \setminus \{\downprob\}$ and we adopt the
convention that $\vtrans^\prime(\downprob)$ is the left derivative as $u \to \downprob$.
\begin{theorem}\label{prop:copula-vtransform}
Let $V$ and $U$ be random variables related by the v-transform $V=\vtrans(U)$. 
\begin{enumerate}
\item The joint distribution function of $(U, V)$ is given by the
  copula
  \begin{equation}
    \label{eq:6}
  C(u,v) =  \P\left(U \leq u, V \leq v\right) = 
\begin{cases}
0 & u < \vtrans^{-1}(v) \\
u - \vtrans^{-1}(v) & \vtrans^{-1}(v) \leq u < \vtrans^{-1}(v)+v \\
v & u \geq \vtrans^{-1}(v)+v\,.
\end{cases}
\end{equation}
\item Conditional on $V = v$ the distribution of $U$ is given by
  \begin{equation}\label{eq:35}
    U =
    \begin{cases}
      \vtrans^{-1}(v) & \text{with probability $\Downprob(v)$ if $v
        \neq 0$} \\
      \vtrans^{-1}(v)+v & \text{with probability $1-\Downprob(v)$ if $v
        \neq 0$} \\
      \downprob & \text{if $v=0$}
      \end{cases}
  \end{equation}
where
\begin{equation}
  \label{eq:7}
  \Downprob(v) = - \frac{1}{\vtrans^\prime(\vtrans^{-1}(v))} \,.
\end{equation}
\item $\E\left(\Downprob(V)\right) = \downprob$.
\end{enumerate}
\end{theorem}
\begin{remark}
In the case of the symmetric v-transform $\vtrans(u)=|1-2u|$ the copula in~\eqref{eq:6} takes the form $C(u,v) = \max(\min(u+\frac{v}{2}-\frac{1}{2},v),0)$. We note that this copula is related to a special case of the tent map copula family $C^{\mathcal{T}}_\theta$ in~\citet{bib:remillard-13} by $C(u,v) = u - C^{\mathcal{T}}_1(u,1-v)$.
\end{remark}
For the linear v-transform family the
conditional probability $\Downprob(v)$ in~(\ref{eq:7}) satisfies
$\Downprob(v) = \downprob$. This implies that the value of
$V$ contains no information about whether $U$ is likely to be below or above
the fulcrum; the probability is always the same regardless of $V$. In general this is
not the case and the value of
$V$ does contain information about whether $U$ is large or
small.

Part (2) of Theorem~\ref{prop:copula-vtransform} is the key to
stochastically inverting a v-transform in
the general case. Based on this result we
define the concept of stochastic inversion of a v-transform.
We refer to the function
$\Downprob$ as the \textit{conditional down probability} of $\mathcal{V}$.
\begin{definition}[Stochastic inversion function of a
  v-transform]\label{def:stochinverse}
  Let
$\vtrans$ be a v-transform with conditional down probability
$\Delta$. The two-place function
$\bm{\vtrans}^{-1} : [0,1] \times [0,1] \to [0,1]$ defined by
\begin{equation}\label{eq:1B}
  \bm{\vtrans}^{-1}(v,w) = 
  \begin{cases}
    \vtrans^{-1}(v) & \text{if $w \leq \Delta(v)$} \\
v + \vtrans^{-1}(v) & \text{if $w > \Delta(v)$.}
\end{cases}
\end{equation}
is the stochastic inversion function of $\vtrans$.
\end{definition}

\noindent The following proposition, which generalizes
Lemma~\ref{lemma:reconstructU}, allows us to construct
general asymmetric processes that generalize the process of Algorithm~\ref{algo1}.
\begin{proposition}\label{prop:resconstructU}
Let $V$ and $W$ be iid $U(0,1)$ variables and
let $\vtrans$ be a v-transform with stochastic inversion function $\bm{\vtrans}$. If $U =
\bm{\vtrans}^{-1}(V,W)$, then  $\vtrans(U) = V$ and $U \sim U(0,1)$.
\end{proposition}



In Section~\ref{sec:properties-model} we apply v-transforms and their stochastic inverses to the terms of
time series models. To understand the effect this has on the serial
dependencies between random variables, we need to consider
multivariate componentwise v-transforms of random vectors with uniform
marginal distributions and
these can also be represented in terms of copulas.
We now give a result which forms the basis for the analysis of serial
dependence properties. The first part of the result shows the
relationship between copula densities under componentwise
v-transforms. The second part shows the relationship under the
componentwise stochastic inversion of a v-transform; in this case we
assume that the stochastic inversion of each term takes place
independently given $\bm{V}$ so that all serial dependence comes from
$\bm{V}$. 

\begin{theorem}\label{theorem:multivariate-vtransform}
Let $\vtrans$ be a v-transform and let $\bm{U} =
(U_1,\ldots,U_d)^\prime$ and $\bm{V} = (V_1,\ldots,V_d)^\prime$ be
vectors of uniform random variables with copula densities $c_{\bm{U}}$
and $c_{\bm{V}}$ respectively.
\begin{enumerate}
\item If $\bm{V} = (\vtrans(U_1),\ldots,\vtrans(U_d))^\prime$ then
\begin{equation}\label{eq:16}
c_{\bm{V}}(v_1, \ldots, v_d) = \sum_{j_1=1}^2 \cdots \sum_{j_d=1}^2
c_{\bm{U}}(u_{1j_1}, \ldots,u_{d j_d}) \prod_{i=1}^d
\Delta(v_i)^{\indicator{j_i=1}} \left(1- \Delta(v_i)\right)^{\indicator{j_i = 2}}
\end{equation}
where $u_{i1} =  \vtrans^{-1}(v_i)$ and $u_{i2} =  \vtrans^{-1}(v_i) + v_i$ for all $i\in\{1,\ldots,d\}$.
\item
If $\bm{U} =
(\bm{\vtrans}^{-1}(V_1 , W_1),\ldots,\bm{\vtrans}^{-1}(V_d, W_d))^\prime$ where
$W_1,\ldots,W_d$ are iid uniform random variables that are also
independent of
$V_1,\ldots,V_d$, then
\begin{equation}\label{eq:15}
  c_{\bm{U}}(u_1,\ldots,u_d) = c_{\bm{V}}(\vtrans(u_1),\ldots,\vtrans(u_d)).
\end{equation}
\end{enumerate}
\end{theorem}

\section{VT-ARMA copula models}\label{sec:properties-model}

In this section we study some properties of the class of time series models obtained by the following algorithm, which generalizes Algorithm~\ref{algo1}. The models obtained are described as VT-ARMA processes since they are stationary time series constructed using the fundamental building blocks of a v-transform $\vtrans$ and an ARMA process.
\begin{framed}
\begin{algorithm}\label{algo2}
\begin{enumerate}
\item Generate $(Z_t)$ as a causal and invertible Gaussian ARMA process of order $(p,q)$
  with mean zero and variance one.
\item Form the volatility PIT process $(V_t)$ where $V_t =\Phi(Z_t)$
  for all $t$.
\item Generate iid $U(0,1)$ random variables $(W_t)$.
\item Form the series PIT process $(U_t)$ by taking the stochastic
  inverses $U_t = \bm{\vtrans}^{-1}(V_t, W_t)$.
\item Form the process $(X_t)$ by setting $X_t = F_X^{-1}(U_t)$ for some continuous cdf $F_X$.
\end{enumerate}
\end{algorithm}
\end{framed}
We can add any marginal behaviour in the final step and this allows for an infinitely rich choice. We can, for instance, even impose an
infinite-variance or an infinite-mean distribution, such
as the Cauchy distribution, and still obtain a strictly stationary process for $(X_t)$. We make the following definitions.
\begin{definition}[VT-ARMA and VT-ARMA copula process]\label{def:svpit-process}
Any stochastic process $(X_t)$ that can be generated using Algorithm~\ref{algo2} by choosing an underlying ARMA
process with mean zero and variance one, a v-transform $\vtrans$ and and a continuous distribution function $F_X$ is a VT-ARMA process. The process $(U_t)$ obtained at the penultimate step of the algorithm is a VT-ARMA copula process.
\end{definition}

 Figure~\ref{fig:3} gives an example of a simulated process using Algorithm~\ref{algo2} and the v-transform $\vtrans_{\downprob,\kappa,\xi}$ in~\eqref{eq:11} with $\kappa=0.9$ and MA parameter $\xi =1.1$.
 The marginal distribution is a heavy-tailed skewed Student distribution of type~\eqref{eq:10} with degrees-of-freedom $\nu=3$ and skewness 
 $\gamma=0.8$, which gives rise to more large negative returns than large positive returns. The underlying time series model is an ARMA(1,1) model with AR parameter $\alpha=0.95$ and MA parameter $\beta =-0.85$. See caption of figure for full details of parameters.

In the remainder of this section we concentrate on the properties of VT-ARMA copula processes $(U_t)$ from which related properties of VT-ARMA processes $(X_t)$ may be easily inferred.

\subsection{Stationary distribution}\label{sec:uncond-distr}

The VT-ARMA copula process $(U_t)$ of Definition~\ref{def:svpit-process} is a strictly
stationary process since the joint distribution of
$(U_{t_1},\ldots,U_{t_k})$ for
any set of indices $t_1< \cdots < t_k$ is invariant under time shifts.
This property follows easily from the strict stationarity of the underlying
ARMA process $(Z_t)$ according to the following result, which uses
Theorem~\ref{theorem:multivariate-vtransform}.

\begin{proposition}\label{theorem:uncond-copula}
Let $(U_t)$ follow a VT-ARMA copula process with
v-transform $\vtrans$ and an underlying
ARMA($p$,$q$) structure with autocorrelation function $\rho(k)$. The
random vector $(U_{t_1},\ldots,U_{t_k})$ for $k \in \N$ has joint density
$c^{\text{Ga}}_{P(t_1,\ldots,t_k)}(\vtrans(u_{1}),\ldots,\vtrans(u_{k}))$ where
$c^{\text{Ga}}_{P(t_1,\ldots,t_k)}$ denotes the density of the Gaussian copula $C^{\text{Ga}}_{P(t_1,\ldots,t_k)}$ and
$P(t_1,\dots,t_k)$ is a correlation matrix with $(i,j)$ element given by $\rho(| t_j - t_i|)$.
\end{proposition}

An expression for the joint density facilitates the calculation of a number of
dependence measures for the bivariate marginal distribution of $(U_t,
U_{t+k})$. In the bivariate case the correlation matrix of the underlying Gaussian copula $C^{\text{Ga}}_{P(t,t+k)}$ contains a single off-diagonal value $\rho(k)$ and we simply write $C^{\text{Ga}}_{\rho(k)}$. The Pearson correlation of $(U_t, U_{t+k})$ is given by
\begin{align}\label{eq:29}
\rho(U_t, U_{t+k}) &= 12 \int_0^1 \int_0^1 u_1 u_2 c^{\text{Ga}}_{\rho(k)}\left(\vtrans(u_1),
  \vtrans(u_2)\right) \rd u_1 \rd u_2  -3 \;.
\end{align}
This value is also the value of the Spearman rank correlation $\rho_S(X_t, X_{t+k})$ for a VT-ARMA process $(X_t)$ with copula process $(U_t)$ (since the Spearman's rank correlation of a pair of continuous random variables is the Pearson correlation of their copula).
The calculation of~\eqref{eq:29} typically
requires numerical integration. However, in the
special case of the linear v-transform $\vtrans_\downprob$
in~\eqref{eq:1} we can get a simpler expression
as shown in the following result.
\begin{proposition}\label{prop:ARMA-dependence}
Let $(U_t)$ be a VT-ARMA copula process satisfying the assumptions of
Proposition~\ref{theorem:uncond-copula} with linear v-transform $\vtrans_\downprob$. Let $(Z_t)$ denote the
underlying Gaussian ARMA process. Then
  \begin{eqnarray}
    \rho(U_t, U_{t+k}) & = &  (2\downprob-1)^2\rho_S(Z_t,Z_{t+k}) = \frac{6 (2\downprob-1)^2
                               \arcsin\left(\frac{\rho(k)}{2}\right)}{\pi}\;.
                             \label{eq:31}
  \end{eqnarray}
\end{proposition}

For the symmetric v-transform $\vtrans_{0.5}$, equation~\eqref{eq:31} obviously yields a correlation of zero so that, in this case, the VT-ARMA copula process $(U_t)$ is a white noise with an autocorrelation function that is zero, except at lag zero. However even a very asymmetric model
with $\downprob=0.4$ or $\downprob=0.6$ gives $ \rho(U_t, U_{t+k}) =0.04  \rho_S(Z_t,
Z_{t+k})$ so that serial correlations tend to be very weak.

When we add a marginal distribution, the resulting process $(X_t)$ has a different auto-correlation function to $(U_t)$, but the same rank autocorrelation function. 
The symmetric model of
Section~\ref{sec:simple-symm-model} is a white noise process. General
asymmetric processes $(X_t)$ are not perfect white noise processes but
have only very weak serial correlation.


\subsection{Conditional distribution}\label{sec:cond-distr}

To derive the conditional distribution of a VT-ARMA copula process we use the vector notation $\bm{U}_t = (U_1,\ldots,U_t)^\prime$ and $\bm{Z}_t =
  (Z_1,\ldots,Z_t)^\prime$ to denote the history of processes up to
  time point $t$ and $\bm{u}_t$ and $\bm{z}_t$ for realizations. These
  vectors are related by the componentwise transformation $\bm{Z}_{t}=
  \Phi^{-1}(\vtrans(\bm{U}_{t})) $. We assume all processes have
  time index set given by $t \in \{1,2,\ldots\}$.

\begin{proposition}\label{theorem:cond-density}
For $t > 1$ the conditional density $f_{U_t \mid \bm{U}_{t-1}}(u \mid \bm{u}_{t-1})$ is
given by 
\begin{equation}
f_{U_t \mid \bm{U}_{t-1}}(u \mid \bm{u}_{t-1}) =
\frac{\phi\left(
\frac{\Phi^{-1}\left(\vtrans(u)\right) -\mu_t}{\sigma_\epsilon}
\right)}{\sigma_\epsilon \phi\left(\Phi^{-1}(\vtrans(u))\right)}
                                  \label{eq:22}
                                \end{equation}
                              where  $\mu_t= \E(Z_t \mid
\bm{Z}_{t-1} = \Phi^{-1}(\vtrans(\bm{u}_{t-1})) )$  and $\sigma_\epsilon$ is the standard
deviation of the innovation process for the ARMA model followed by $(Z_t)$.
\end{proposition}
When $(Z_t)$ is iid white
noise $\mu_t = 0$,
$\sigma_\epsilon = 1$ and ~\eqref{eq:22} reduces to the uniform density
$f_{U_t \mid \bm{U}_{t-1}}(u \mid \bm{u}_{t-1}) = 1$ as expected.
In the case of the first-order Markov AR(1) model $Z_t = \alpha_1 Z_{t-1} + \epsilon_t$ the conditional mean of $Z_t$ is $\mu_t = \alpha_1 \Phi^{-1}\left(\vtrans(u_{t-1})\right)$ and $\sigma_\epsilon^2 = 1-\alpha_1^2$. The conditional density~\eqref{eq:22} can be easily shown to simplify to
$f_{U_t \mid U_{t-1}}(u \mid u_{t-1}) = c_{\alpha_1}^{\text{Ga}}\left(\vtrans\left(u\right),\vtrans\left(u_{t-1}\right)\right)$
where $c_{\alpha_1}^{\text{Ga}}\left(\vtrans\left(u_1\right),\vtrans\left(u_2\right)\right)$ denotes the copula density derived in Proposition~\ref{theorem:uncond-copula}.
In this special case the VT-ARMA model falls within the class of
first-order Markov copula models considered
by~\citet{bib:chen-fan-06b}, although the copula is new. 

\begin{figure}[htb]
  \centering
   \includegraphics[width=16cm,height=12cm]{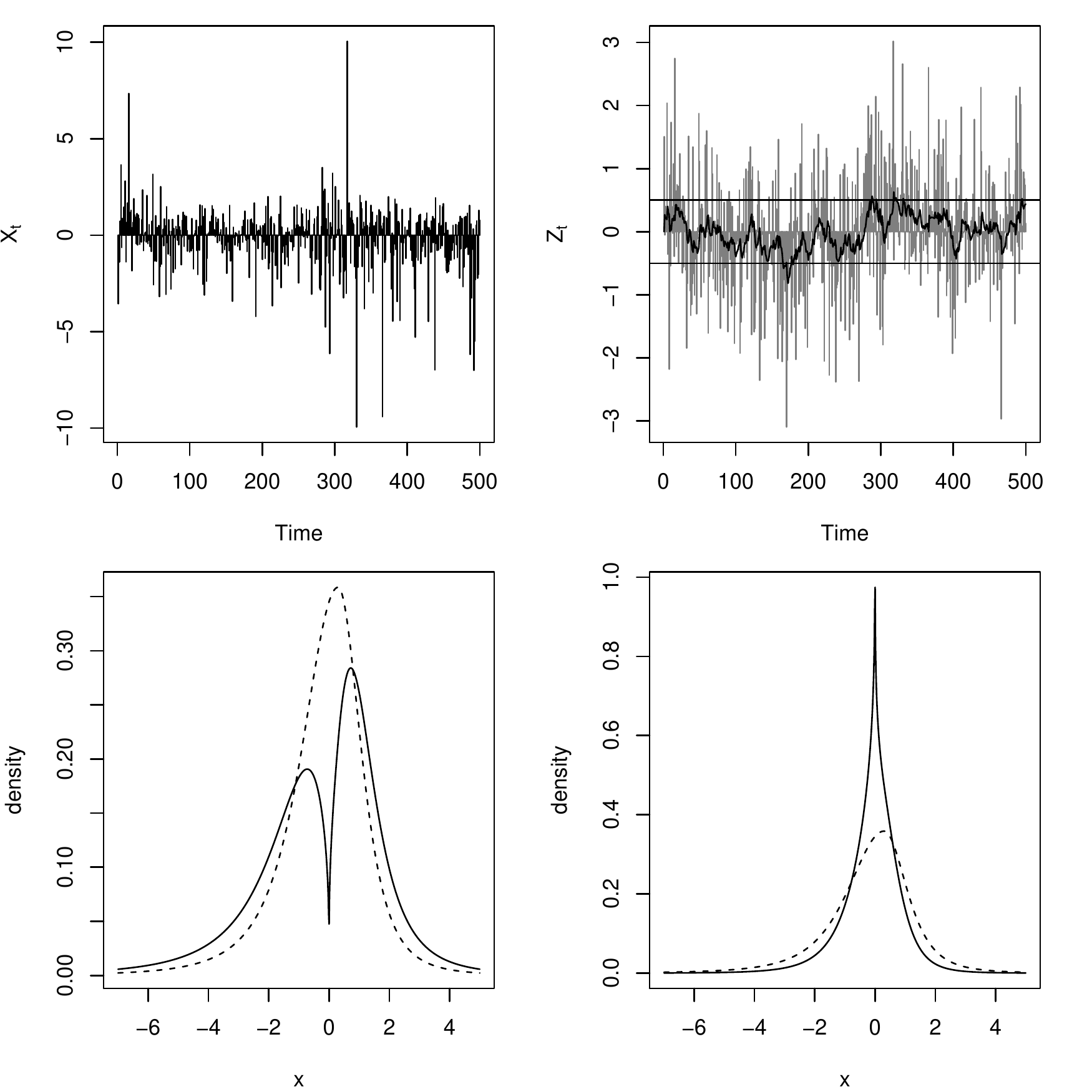}
   \caption{\label{fig:3} Top left: realization of length $n=500$ of $(X_t)$ for a process with a marginal skewed Student distribution (parameters: $\nu=3$, $\gamma=0.8$, $\mu=0.3$, $\sigma=1$) a v-transform of the form~\eqref{eq:11} (parameters: $\downprob=0.50$, $\kappa=0.9$, $\xi=1.1$) and an underlying ARMA process ($\alpha=0.95$, $\beta =-0.85$, $\sigma_\epsilon =0.95$). Top right: the underlying ARMA process $(Z_t)$ in gray with the conditional mean $(\mu_t)$ superimposed in black; horizontal lines at $\mu_t = 0.5$ (a high value) and $\mu_t = -0.5$ (a low value). The corresponding conditional densities are shown in the bottom figures  with the marginal density as a dashed line.}
 \end{figure}

If we add a marginal distribution $F_X$ to the VT-ARMA copula model to obtain a model for $(X_t)$ and use similar notational conventions as above, the resulting VT-ARMA model has conditional density
\begin{equation}\label{eq:501}
f_{X_t \mid \bm{X}_{t-1}}(x \mid \bm{x}_{t-1}) = f_X(x) f_{U_t \mid \bm{U}_{t-1}}(F_X(x) \mid F_X(\bm{x}_{t-1}))
\end{equation}
with $f_{U_t \mid \bm{U}_{t-1}}$ as in~\eqref{eq:22}.
An interesting property of the VT-ARMA process is that the conditional density~\eqref{eq:501} can have a pronounced bimodality for values of $\mu_t$ in excess of zero, that is in high volatility situations where the conditional mean of $Z_t$ is higher than the marginal mean value of zero; in low volatility situations the conditional density appears more concentrated around zero. This phenomenon is illustrated in Figure~\ref{fig:3}. The bimodality in high volatility situations makes sense: in such cases it is likely that the next return will be large in absolute value and relatively less likely that it will be close to zero.

The conditional distribution function of $(X_t)$ is
$F_{X_t \mid \bm{X}_{t-1}}(x \mid \bm{x}_{t-1}) = F_{U_t \mid \bm{U}_{t-1}}(F_X(x) \mid F_X(\bm{x}_{t-1}))$
and hence the $\psi$-quantile $x_{\psi,t}$ of  $F_{X_t\mid\bm{X}_{t-1}}$ can be obtained by
        solving
        \begin{equation}\label{eq:VaR}
          \psi = F_{U_t\mid\bm{U}_{t-1}}(F_X(x_{\psi,t})\mid F_X(\bm{x}_{t-1}))\,.
        \end{equation}
For $\psi < 0.5$ the negative of this value is often referred to as the
    conditional $(1-\psi)$-VaR (value-at-risk) at time $t$ in
    financial applications.

\section{Statistical inference}\label{sec:estimation-model}

In the copula approach to dependence modelling, the copula is the
object of central interest and marginal distributions are often of
secondary importance. A number of different approaches to estimation
are found in the literature. As before, let $x_1,\ldots,x_n$ represent realizations of variables $X_1,\ldots,X_n$ from the time series process $(X_t)$.

The semi-parametric
  approach developed by~\citet{bib:genest-ghoudi-rivest-95} is very
  widely used in copula inference and has been applied
  by~\citet{bib:chen-fan-06b} to first-order Markov copula models in the time series context. In this approach the marginal distribution
  $F_X$ is first estimated non-parametrically using
  the scaled empirical distribution function $F_n^{(X)}$ (see definition in Section~\ref{sec:intro}) and the data are transformed onto the $(0,1)$ scale. This has the effect of creating pseudo-copula data $u_t = \text{rank}(x_t)/(n+1)$ where $\text{rank}(x_t)$ denotes the rank of $x_t$ within the sample. The copula is fitted to the pseudo-copula data by maximum likelihood (ML). 
  
As an alternative, the inference-functions-for-margins (IFM) approach
of~\citet{bib:joe-15} could be applied. This is also a two-step method although in this case
a parametric model $\widehat{F}_X$ is estimated under an iid
assumption in the first step and the
copula is fitted to the data $u_t = \widehat{F}_X(x_t)$ in the second
step.

The approach we adopt for our empirical example is to first use the
semi-parametric approach to determine a reasonable copula process, then to
estimate marginal parameters under an iid assumption, and finally to
estimate all parameters jointly using the parameter estimates from the
previous steps as starting values.

We concentrate on the mechanics of deriving maximum likelihood
estimates (MLEs). The problem of establishing the asymptotic properties
of the MLEs in our setting is a difficult one. It is similar to, but
appears to be more
technically challenging than, the problem of showing consistency and efficiency
of MLEs for a Box-Cox-transformed Gaussian ARMA process, as discussed
in~\citet{bib:terasaka-hosoya-07}.
We are also working with a componentwise transformed ARMA process, although in our
case the transformation $(X_t) \to (Z_t)$ is via the non-linear,
non-increasing volatility proxy transformation $T^{(Z)}(x)$ in~(\ref{eq:volproxyTz}), which is not differentiable at
the change point $\mu_T$. We have, however, run extensive simulations
which suggest good behaviour of the MLEs in large samples.

\subsection{Maximum likelihood estimation of VT-ARMA copula process}\label{sec:MLestimation}

We first consider the estimation of the VT-ARMA copula process
for a sample of data $u_1,\ldots,u_n$.
Let $\bm{\theta}^{(V)}$ and $\bm{\theta}^{(A)}$ denote the parameters of the v-transform and ARMA
model respectively. It follows from Theorem~\ref{theorem:multivariate-vtransform} (part 2) and
Proposition~\ref{theorem:uncond-copula} that the log-likelihood for
the sample $u_1,\ldots,u_n$ is simply the log density of the Gaussian
copula under componentwise inverse v-transformation. This is given by
\begin{equation}\label{eq:loglik}
\begin{split}
  L (\bm{\theta}^{(V)}, \bm{\theta}^{(A)} \mid u_1,\ldots,u_n) &=
L^*(\bm{\theta}^{(A)}\mid \Phi^{-1}(\vtrans_{\bm{\theta}^{(V)}}(u_1)),\ldots,
                                        \Phi^{-1}(\vtrans_{\bm{\theta}^{(V)}}(u_n))) \\
& \hspace{6cm} - \sum_{t=1}^n \ln \phi\left(\Phi^{-1}\left( \vtrans_{\bm{\theta}^{(V)}}(u_t) \right) \right)
\end{split}
\end{equation}
where the first term $L^*$ is the
                                       log-likelihood for an ARMA
                                       model with a standard N(0,1)
                                       marginal distribution. Both terms in the log-likelihood~\eqref{eq:loglik} are relatively straightforward to evaluate.

The evaluation of the ARMA likelihood $L^*(\bm{\theta}^{(A)}
                                       \mid z_1,\ldots,z_n) $ for
                                       parameters $\bm{\theta}^{(A)}$
                                       and data $z_1,\ldots,z_n$ can
                                       be accomplished using the
                                       Kalman filter.
However, it is important to note that the assumption that the data $z_1,\ldots,z_n$ are standard 
                   normal requires a bespoke
                                       implementation of the Kalman
                                       filter, since standard
                                       software always treats the
                                       error variance $\sigma^2_\epsilon$ as a free
                                       parameter in the ARMA model.
In our case we need to constrain $\sigma^2_\epsilon$ to be a function
of the ARMA parameters so that
$\var(Z_t) =1$. For example, in the case of an
ARMA(1,1) model with AR parameter $\alpha_1$ and MA parameter
$\beta_1$, this means that
  $\sigma_\epsilon^2 = \sigma_\epsilon^{2} (\alpha_1,\beta_1)  =
  (1-\alpha_1^2)/(1 + 2\alpha_1\beta_1 + \beta_1^2)$.
The constraint on $\sigma^2_\epsilon$ must be incorporated into the state-space
representation of the ARMA model.


Model validation tests for the VT-ARMA copula can be based on residuals
\begin{equation}\label{eq:residuals}
r_t = z_t - \widehat{\mu}_t,\quad z_t = \Phi^{-1}(\vtrans_{\widehat{\bm{\theta}}^{(V)}}(u_t)))
\end{equation}
where $z_t$ denotes the implied realization of the normalized
volatility proxy variable and where an estimate  $\widehat{\mu}_t$ of
the conditional mean $\mu_t = \E(Z_t \mid \bm{Z}_{t-1}=\bm{z}_t)$ may
be obtained as an output of the Kalman filter. The residuals should
behave like an iid sample from a normal distribution.

Using the estimated model, it is also possible to implement a
likelihood-ratio (LR) test for the presence of stochastic volatility
in the data. Under the null hypothesis that $\bm{\theta}^{(A)} =
\bm{0}$
the log-likelihood~(\ref{eq:loglik}) is identically equal to zero. Thus the size
of the maximized log-likelihood $L(\widehat{\bm{\theta}}^{(V)},
\widehat{\bm{\theta}}^{(A)}\,;\, u_1,\ldots,u_n)$ provides a measure
of the evidence for the presence of stochastic volatility.

\subsection{Adding a marginal model}
If  $F_X$ and $f_X$ denote the cdf and density of the marginal model
and the parameters are denoted $\bm{\theta}^{(M)}$ then the full
log-likelihood for the data $x_1,\ldots,x_n$ is simply
\begin{multline}\label{eq:likelihood-step1}
L^{\text{full}}(\bm{\theta} \mid x_1,\ldots,x_n) = \sum_{t=1}^n \ln f_X(x_t \,;\,
                                     \bm{\theta}^{(M)}) \\
                                     +
                                     L\left(\bm{\theta}^{(V)},\bm{\theta}^{(A)} \mid F_X(x_1  \,;\,
                                     \bm{\theta}^{(M)}),\ldots,  F_X(x_n  \,;\,
                                     \bm{\theta}^{(M)}) \right)
\end{multline}
where the first term
is the log-likelihood for a sample of iid data from the marginal
distribution $F_X$ and the second term is~\eqref{eq:loglik}.


When a marginal model is added we can recover the implied form of the
volatility proxy transformation using
Proposition~\ref{prop:reconstructT}. If $\widehat{\downprob}$ is the
estimated fulcrum parameter of the v-transform then the estimated
change point is $ \widehat{\mu}_T =F_X^{-1}(\widehat{\downprob}; \widehat{\bm{\theta}}^{(M)})$ and the implied profile function is
   \begin{eqnarray}~\label{eq:proxy-profile}
\widehat{g}_T(x) & = & \widehat{F}_X^{-1}\left( \widehat{F}_X(\widehat{\mu}_T-x) - \vtrans_{\widehat{\bm{\theta}}^{(V)}}\left(  \widehat{F}_X(\widehat{\mu}_T-x) \right)\right) - \widehat{\mu}_T.
      \end{eqnarray}
 Note that is is possible to force the change point to be zero in a
 joint estimation of marginal model and copula by imposing the
 constraint $F_X(0; \bm{\theta}^{(M)}) = \downprob$ on the fulcrum and
 marginal parameters during the optimization. However, in our
 experience, superior fits are obtained when these parameters are unconstrained.

\subsection{Example}\label{sec:examples}

We analyse $n=1043$ daily log-returns for the Bitcoin price series for
the period 2016--2019; values are multiplied by 100. 
We first apply the semi-parametric
  approach of~\citet{bib:genest-ghoudi-rivest-95} using the
  log-likelihood~(\ref{eq:loglik}) which yields the results in
  Table~\ref{table1}. Different models are referred to by VT($n$)-ARMA($p$, $q$) where $(p,q)$ refers to the ARMA model and $n$ indexes the
v-transform: 1 is the linear v-transform $\vtrans_\downprob$
in~\eqref{eq:1}; 3 is the three-parameter transform
$\vtrans_{\downprob,\kappa,\xi}$ in~\eqref{eq:11}; 2 is the two-parameter v-transform given by
$\vtrans_{\downprob,\kappa}:= \vtrans_{\downprob,\kappa,1}$.
In unreported analyses we also tried the three-parameter family based on the
beta distribution, but this had negligible effect on the results.

The column marked $L$ gives the value of the maximized
log-likelihood. All values are large and positive showing strong
evidence of stochastic volatility in all cases. The model
VT(1)-ARMA(1,0) is a first-order Markov model with linear
v-transform. The fit of this model is noticeably poorer than the
others suggesting that Markov models are insufficient to capture the
persistence of stochastic volatility in the data.
The column marked SW contains the p-value for a Shapiro-Wilks test of
normality applied to the residuals from the VT-ARMA copula model; the
result is non-significant in all cases.


\begin{table}[ht]
\centering
\begingroup\setlength{\tabcolsep}{4pt}
\begin{tabular}{lrrrrrrrr}
  \toprule
Model & $\alpha_1$ & $\beta_1$ & $\delta$ & $\kappa$ & $\xi$ & SW & $L$ & AIC \\ 
  \midrule
VT(1)-ARMA(1,0) & 0.283 &  & 0.460 &  &  & 0.515 & 37.59 & -71.17 \\ 
   & 0.026 &  & 0.001 &  &  &  &  &  \\ 
  VT(1)-ARMA(1,1) & 0.962 & -0.840 & 0.416 &  &  & 0.197 & 92.91 & -179.81 \\ 
   & 0.012 & 0.028 & 0.004 &  &  &  &  &  \\ 
  VT(2)-ARMA(1,1) & 0.965 & -0.847 & 0.463 & 0.920 &  & 0.385 & 94.73 & -181.45 \\ 
   & 0.011 & 0.026 & 0.001 & 0.131 &  &  &  &  \\ 
  VT(3)-ARMA(1,1) & 0.962 & -0.839 & 0.463 & 0.881 & 0.995 & 0.407 & 94.82 & -179.64 \\ 
   & 0.012 & 0.028 & 0.001 & 0.123 & 0.154 &  &  &  \\ 
   \bottomrule
\end{tabular}
\endgroup
\caption{Analysis of daily Bitcoin return data 2016--2019. Parameter estimates, 
              standard errors (below estimates) and information about the fit: 
              SW denotes Shapiro Wilks p-value; 
              $L$ is the maximized value of the log-likelihood and AIC is the Akaike information criterion.} 
\label{table1}
\end{table}

\begin{figure}[htb]
  \centering
   \includegraphics[width=16cm,height=12cm]{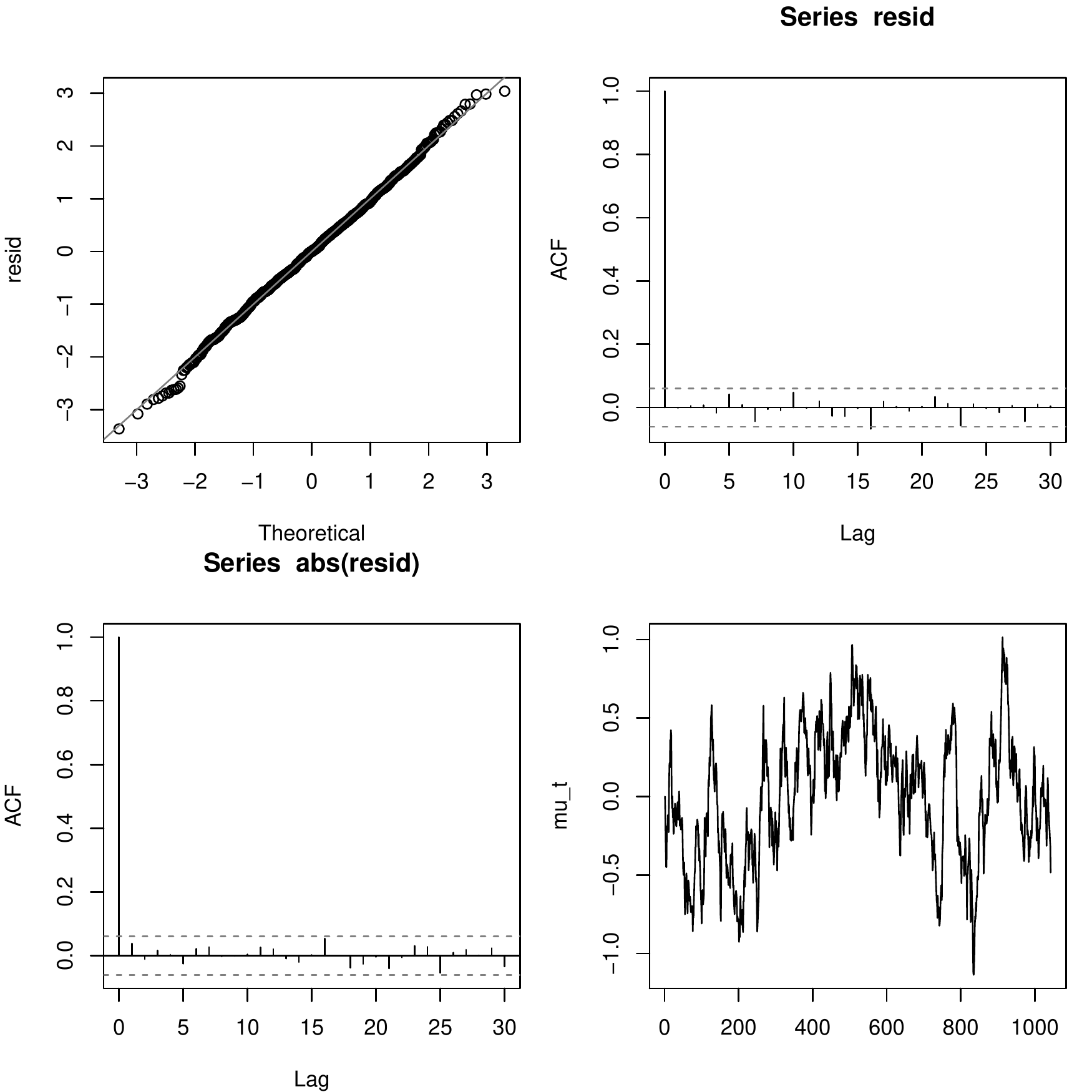}
   \caption{\label{fig:6} Plots for a VT(2)-ARMA(1,1) model fitted to the Bitcoin return data: QQplot of the residuals against normal (upper left); acf of the residuals (upper right); acf of the absolute residuals (lower left);  estimated conditional mean process $(\mu_t)$ (lower right).}
 \end{figure}

According to the AIC values, the VT(2)-ARMA(1,1) is the best model. We experimented with
higher order ARMA processes but this did not lead to further
significant improvements. Figure~\ref{fig:6} provides a visual of the fit of this model. The pictures in the panels show the QQplot of the residuals against normal, acf plots of the residuals and squared residuals and the estimated
conditional mean process $(\widehat{\mu}_t)$, which can be taken as an
indicator of high and low volatility periods. 
The residuals and absolute residuals show very little evidence of
serial correlation and the QQplot is relatively linear, suggesting that the ARMA filter has been successful
in explaining much of the serial dependence structure of the
normalized volatility proxy process.

We  now add various marginal distributions to the VT(2)-ARMA(1,1) copula model and estimate all parameters of the model jointly.
We have experimented with a number of location-scale families
including Student t, Laplace (double exponential) and a double-Weibull
family which generalizes the Laplace distribution and is
constructed by taking back-to-back Weibull distributions. Estimation results are
presented for these 3 distributions in Table~\ref{table2}. All three marginal distributions are symmetric around their location
parameters $\mu$ and no improvement is obtained by adding skewness
using the construction of~\citet{bib:fernandez-steel-98}
described in Section~\ref{sec:flex-param-family}; in fact, the Bitcoin
returns in this time period show a remarkable degree of symmetry. In the table the shape
and scale parameters of the distributions are denoted $\eta$ and $\sigma$
respectively; in the case of Student, an infinite-variance distribution
with degree-of-freedom parameter $\eta = 1.94$ is fitted, but this
model is inferior to the models with Laplace and double-Weibull
margins; the latter is the favoured model on the basis of AIC values.

\begin{table}[ht]
\centering
\begingroup\setlength{\tabcolsep}{4pt}
\begin{tabular}{lrrrrrr}
  \toprule
 & Student &  & Laplace &  & dWeibull &  \\ 
  \midrule
$\alpha_1$ & 0.954 & 0.012 & 0.953 & 0.012 & 0.965 & 0.021 \\ 
  $\beta_1$ & -0.842 & 0.026 & -0.847 & 0.025 & -0.847 & 0.035 \\ 
  $\delta$ & 0.478 & 0.001 & 0.480 & 0.002 & 0.463 & 0.000 \\ 
  $\kappa$ & 0.790 & 0.118 & 0.811 & 0.129 & 0.939 & 0.138 \\ 
  $\eta$ & 1.941 & 0.005 &  &  & 0.844 & 0.022 \\ 
  $\mu$ & 0.319 & 0.002 & 0.315 & 0.002 & 0.192 & 0.001 \\ 
  $\sigma$ & 2.427 & 0.003 & 3.194 & 0.004 & 2.803 & 0.214 \\ 
   \midrule
SW & 0.585 &  & 0.551 &  & 0.376 &  \\ 
  $L$ & -2801.696 &  & -2791.999 &  & -2779.950 &  \\ 
  AIC & 5617.392 &  & 5595.999 &  & 5573.899 &  \\ 
   \bottomrule
\end{tabular}
\endgroup
\caption{VT(2)-ARMA(1,1) model with 3 different margins: Student t, Laplace, double Weibull. 
              Parameter estimates, standard errors (alongside estimates) and information about the fit: 
              SW denotes Shapiro Wilks p-value; $L$ is the maximized value of the log-likelihood and 
              AIC is the Akaike information criterion.} 
\label{table2}
\end{table}

\begin{figure}[htb]
  \centering
   \includegraphics[width=16cm,height=12cm]{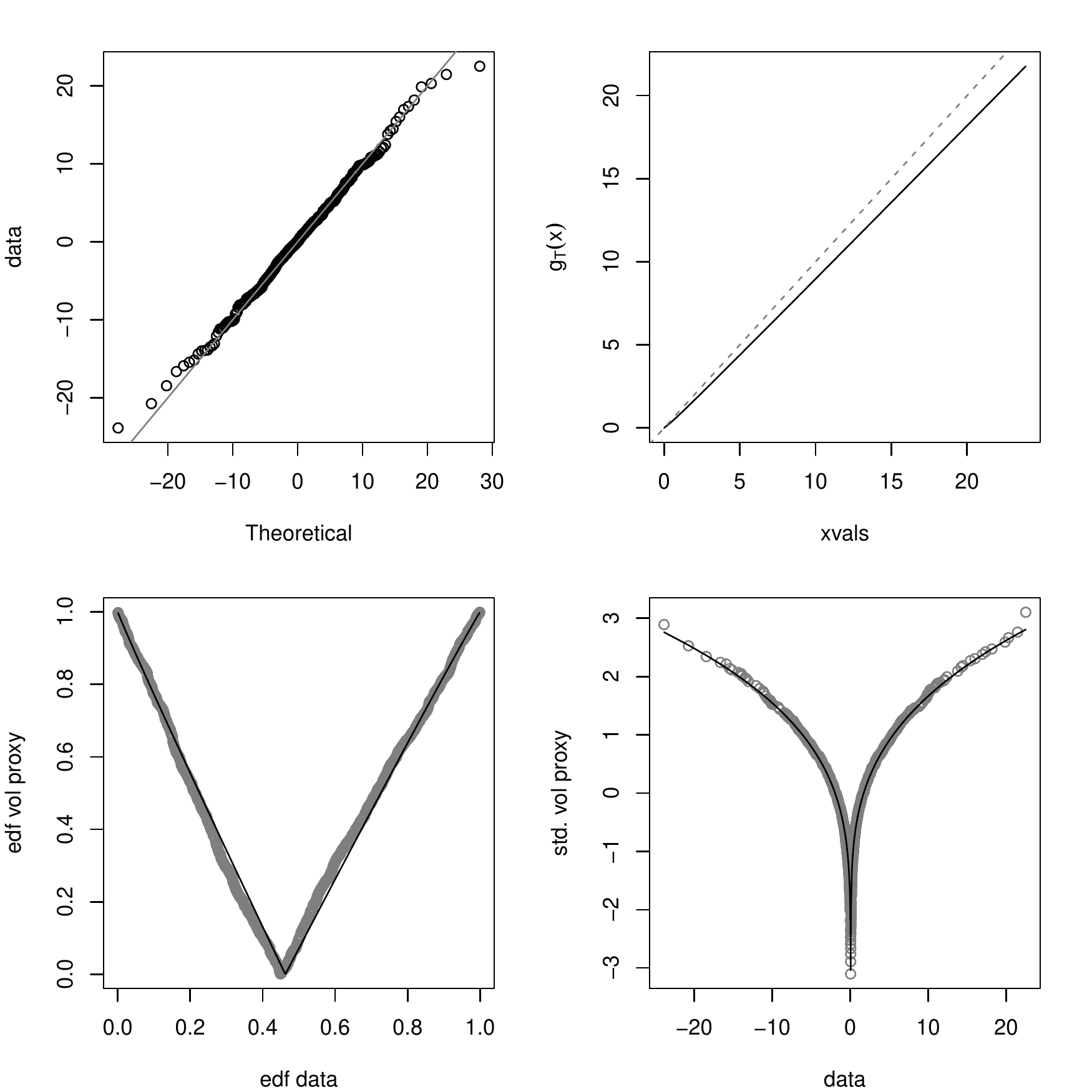}
   \caption{\label{fig:7} Plots for a VT(2)-ARMA(1,1) model combined
     with a double Weibull marginal distribution fitted to the Bitcoin
     return data: QQplot of the data against fitted double Weibull model (upper left); estimated volatility proxy profile function $g_T$ (upper right); estimated v-transform (lower left);  implied relationship between data and volatility proxy variable (lower right).}
 \end{figure}

Figure~\ref{fig:7} shows some aspects of the joint fit for the fully
parametric VT(2)-ARMA(1,1) model with double-Weibull margin. A QQplot of the data against the
fitted marginal distribution confirms that the double-Weibull is a good
marginal model for these data. Although this distribution is
sub-exponential (heavier-tailed than exponential), its tails do not follow
a power law and it is in the maximum domain of attraction of the
Gumbel distribution~\citep[see, for example,][Chapter 5]{bib:mcneil-frey-embrechts-15}.

Using~\eqref{eq:proxy-profile} the implied volatility
proxy profile function $\widehat{g}_T$ can be constructed and is found
to lie just below the line $y=x$ as shown in the upper-right
panel. The change point is estimated to be $ \widehat{\mu}_T = 0.06$. 
We can also estimate an implied
volatility proxy transformation in the equivalence
class defined by $\widehat{g}_T$
and $\widehat{\mu}_T$. We estimate the transformation
$T = T^{(Z)}$ in~(\ref{eq:volproxyTz}) by taking $\widehat{T}(x) = \Phi^{-1}( \vtrans_{\widehat{\bm{\theta}}^{(V)}}(F_X( x ;
\widehat{\bm{\theta}}^{(M)})))$.
In the lower-left panel of Figure~\ref{fig:7} we show the empirical
v-transform formed from the data $(x_t,\widehat{T}(x_t))$ together
with the fitted parametric v-transform
$\vtrans_{\widehat{\bm{\theta}}^{(V)}}$.  We recall from
Section~\ref{sec:intro} that the empirical v-transform is
the plot $(u_t, v_t)$ where $u_t =
 F^{(X)}_n(x_t)$ and $v_t =
 F^{(\widehat{T}(X))}_n(\widehat{T}(x_t))$.
The empirical v-transform and the fitted parametric v-transform show a
good degree of correspondence.
The lower-right panel of Figure~\ref{fig:7} shows the 
volatility proxy transformation $\widehat{T}(x)$ as a
function of $x$ superimposed on the points $(x_t,
\Phi^{-1}(v_t))$. Using the curve we can compare the effects of, for
example, a log-return ($\times$ 100) of -10 and a log-return of
10. For the fitted model these are 1.55 and 1.66 showing that the up
movement is associated with slightly higher volatility.

 As a comparison to the VT-ARMA model we fitted standard GARCH(1,1)
 models using Student t and generalized error distributions for
 the innovations; these are standard choices available in the popular
 \texttt{rugarch} package in \textsf{R}. The generalized error distribution (GED) contains normal
 and Laplace as special cases as well as a model which has similar
 tail behaviour to Weibull; note, however, that by the theory
 of~\cite{bib:mikosch-starica-00} the tails of the marginal
 distribution of the GARCH decay according to a power law in both cases. The results in Table~\ref{table3} show that the
 VT(2)-ARMA(1,1) models with Laplace and double-Weibull marginal
 distributions outperform both GARCH models in terms of AIC values.

\begin{table}[ht]
\centering
\begingroup\setlength{\tabcolsep}{4pt}
\begin{tabular}{lrr}
  \toprule
 & Parameters & AIC \\ 
  \midrule
VT-ARMA (Student) & 7 & 5617.39 \\ 
  VT-ARMA (Laplace) & 6 & 5596.00 \\ 
  VT-ARMA (dWeibull) & 7 & 5573.90 \\ 
  GARCH (Student) & 5 & 5629.02 \\ 
  GARCH (GED) & 5 & 5611.53 \\ 
   \bottomrule
\end{tabular}
\endgroup
\caption{Comparison of three VT(2)-ARMA(1,1) models with different marginal 
              distributions with two GARCH(1,1) models with different innovation distributions.} 
\label{table3}
\end{table}

\begin{figure}[htb]
  \centering
   \includegraphics[width=16cm,height=12cm]{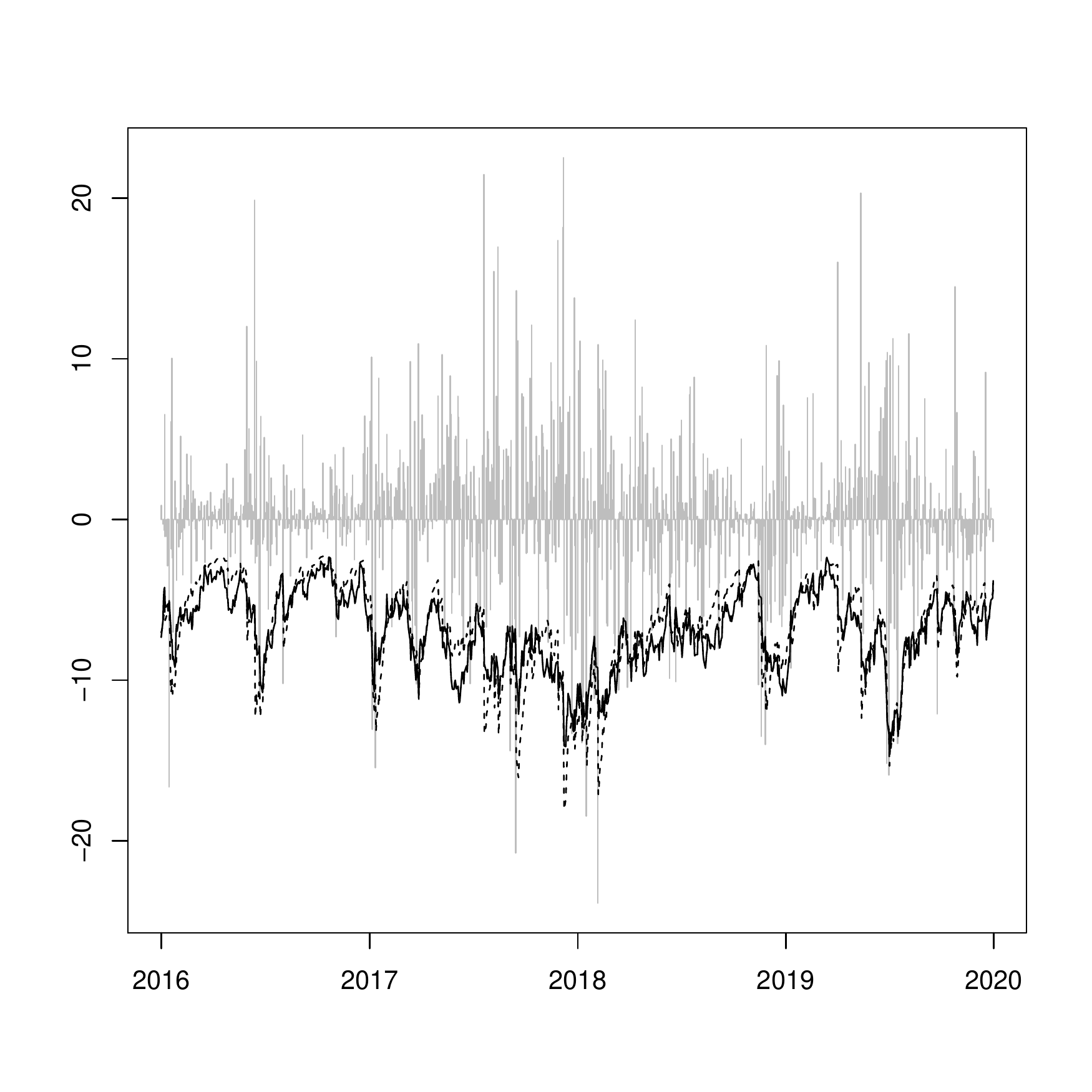}
   \caption{\label{fig:8} Plot of estimated 95\% value-at-risk (VaR)
     for Bitcoin return data superimposed on log returns. Solid line
     shows VaR estimated using the VT(2)-ARMA(1,1) model combined with
     a double Weibull marginal distribution; the dashed line shows VaR estimated using a GARCH(1,1) model with GED innovation distribution.}
 \end{figure}

Figure~\ref{fig:8}
shows the in-sample 95\% conditional value-at-risk (VaR) estimate 
based on the VT(2)-ARMA(1,1) model which has been calculated using~(\ref{eq:VaR}). For comparison, a dashed line shows
the corresponding estimate for the GARCH(1,1) model with GED innovations.

 Finally, we carry out an out-of-sample comparison of
   conditional VaR
  estimates using the same two models. In this analysis, the
  models are estimated daily throughout the 2016--2019 period using a
  1000-day moving data window and one-step-ahead VaR forecasts are
  calculated. The VT-ARMA model gives 47 exceptions of
  the 95\% VaR and 11 exceptions of the 99\% VaR, compared with
  expected numbers of 52 and 10 for a 1043 day sample, while the GARCH model
  leads to 57 and 12 exceptions; both models pass binomial tests for
  these exception counts. In a follow-up paper~\citep{bib:bladt-mcneil-20},
  we conduct more extensive out-of-sample backtests for models using
  v-transforms and copula processes and show that they rival and often
  outperform forecast models from the extended GARCH~family.

 \section{Conclusion}\label{sec:conclusion}
 This paper has proposed a new approach to volatile financial time series 
  in which v-transforms are used to describe the relationship between quantiles
of the return distribution and quantiles of the distribution of a
predictable volatility proxy variable. 
We have characterized v-transforms mathematically and shown that the
stochastic inverse of a v-transform may be used to construct
stationary models for return series where arbitrary marginal
distributions may be coupled with dynamic copula models for the serial dependence in the volatility proxy.

The construction was illustrated using the serial dependence model
implied by a Gaussian ARMA process.  The resulting class of VT-ARMA
processes is able to capture the important features of financial
return series including near-zero
serial correlation (white noise behaviour) and volatility
clustering. Moreover, the
models are relatively straightforward to estimate building
on the classical maximum-likelihood estimation of an ARMA model using
the Kalman filter. This can be accomplished in the stepwise manner
that is typical in copula modelling or through joint modelling of
marginal and copula process. The resulting models yield insights into
the way that volatility responds to returns of different magnitude and
sign and can give estimates of unconditional and conditional quantiles
(VaR) for practical risk measurement purposes. 

There are many possible uses for VT-ARMA
copula processes. Because we have complete control over the marginal
distribution they are very natural candidates for the innovation
distribution in other time series model. For example, they could
be applied to the innovations of an ARMA model to obtain ARMA models
with VT-ARMA errors; this might be particularly appropriate for longer
interval returns, such as weekly or monthly returns, where some serial
dependence is likely to be present in the raw return data. 

Clearly, we could use other copula processes for the volatility PIT
process $(V_t)$. The VT-ARMA copula process has some limitations:
the radial symmetry of the underlying Gaussian copula means
that the serial dependence between large values of the volatility
proxy must mirror the serial dependence between small
values; moreover this copula does not admit tail dependence in either
tail and it seems plausible that very large values of the volatility
proxy might have a tendency to occur in 
succession.

To extend the class of models based on v-transforms we can look for
models for the volatility PIT process $(V_t)$ with higher dimensional
marginal distributions given by asymmetric copulas with
upper tail dependence. First-order Markov copula models as developed
in~\citet{bib:chen-fan-06b} can give asymmetry and tail
dependence, but they cannot model the dependencies at longer lags that we
find in empirical data. D-vine copula models can model higher-order
Markov dependencies and~\cite{bib:bladt-mcneil-20} show that this is a promising alternative
specification for the volatility PIT process.


\section*{Software}

The analyses were carried out using \textsf{R} 4.0.2 (R Core Team,
2020) and the \texttt{tscopula}  package (Alexander J.~McNeil and
Martin Bladt, 2020) available at
\texttt{https://github.com/ajmcneil/tscopula}. The full reproducible
code and the data are available at \texttt{https://github.com/ajmcneil/vtarma}.

\section*{Acknowledgements}

The author is grateful for valuable
    input from a number of researchers including Hansjoerg Albrecher,
    Martin Bladt,
    Val\'{e}rie Chavez-Demoulin, Alexandra Dias, Christian Genest,
    Michael Gordy, Yen Hsiao Lok, Johanna Ne\v{s}lehov\'{a}, Andrew Patton and Ruodu
    Wang. Particular thanks are due to Martin Bladt for providing the
    Bitcoin data and collaborating on the data analysis. The paper was completed while the author was a guest at the
  Forschungsinstitut f\"ur Mathematik (FIM) at ETH Zurich.

\appendix
\numberwithin{equation}{section}
 \section{Proofs}\label{sec:proofs}
\subsection{Proof of Proposition~\ref{prop:model-with-skewness}}\label{sec:proof-model-with-skewness}
We observe that for $x \geq 0$
  \begin{displaymath}
    F_{T(X)}(x) = \P(\mu_T-T_1^{-1}(x) \leq X_t \leq \mu_T + T_2^{-1}(x)) = F_X(\mu_T+ T_2^{-1}(x)) -
    F_X(\mu_T-T_1^{-1}(x)).
  \end{displaymath}
$\{ X_t \leq \mu_T\}  \iff \{U \leq F_X(\mu_T)\}$ and in this case
\begin{align*}
  V =  F_{T(X)}(T(X_t)) =  F_{T(X)}(T_1(\mu_T-X_t)) &= F_X(\mu_T + T_2^{-1}(T_1(\mu_T-X_t))) -
    F_X(X_t) \\
&= F_X\left(\mu_T + g_T\left(\mu_T-F_X^{-1}(U)\right)\right) - U.
\end{align*}
$ \{ X_t > \mu_T\} \iff \{U > F_X(\mu_T)\}$ and in this case
\begin{align*}
  V =  F_{T(X)}(T(X_t)) =  F_{T(X)}(T_2(X_t -\mu_T)) &= F_X(X_t) -
    F_X(\mu_T-T_1^{-1}(T_2( X_t -\mu_T) ) ) \\
&= U - F_X\left(\mu_T- g_T^{-1}\left( F_X^{-1}(U) -\mu_T\right)  \right).
\end{align*}


\subsection{Proof of Proposition~\ref{prop:parametric-family}}\label{prop:proof-parametric-family}
The cumulative distribution
function $F_0(x)$ of the double exponential distribution is equal to
$0.5e^x$ for $x \leq 0$ and $1 - 0.5e^{-x}$ if $x>0$.
It is straightforward to verify that
\begin{displaymath}
  F_X(x;\gamma) = 
\begin{cases}
\downprob e^{\gamma x} & x\leq 0 \\
1 - (1-\downprob)e^{-\frac{x}{\gamma}} & x > 0
\end{cases}
\quad\text{and}\quad
  F_X^{-1}(u;\gamma) = 
\begin{cases}
\frac{1}{\gamma} \ln\left(\frac{u}{\downprob}\right) & u\leq \downprob \\
-\gamma \ln \left( \frac{1-u}{1-\downprob}\right) & u > \downprob\,.
\end{cases}
\end{displaymath}
When $g_T(x) = k x^\xi$ we obtain for $u \leq \downprob$ that
\begin{align*}
\vtrans_{\downprob,\kappa,\xi}(u) = F_X\left(\frac{k}{\gamma^\xi} \left(
  \ln \left(\frac{
  \downprob}{u} \right)^\xi\right) ;\gamma \right) - u
&= 1-u -
  (1-\downprob)\exp\left(- \frac{k}{\gamma^{\xi+1}}\left(-\ln\left(\frac{u}{\downprob}\right)\right)^\xi
                                                         \right)\;.
\end{align*}
For $u > \downprob$ we make a similar calculation.

\subsection{Proof of
  Theorem~\ref{theorem:v-characterization}}\label{proof:v-characterization}
It is easy to check that equation~\eqref{eq:2} fulfills the list of
properties in Lemma~\ref{cor:vtranstions}. We concentrate on showing that a function
that has these properties must be of the form~\eqref{eq:2}.
It helps to consider the picture of a v-transform in Figure~\ref{fig:99}.
Consider the lines $v = 1 - u$ and $v = \downprob - u$ for $u \in
[0,\downprob]$. The areas above the former and below the latter are
shaded gray.

The left branch of the v-transform must start at
$(0,1)$, end at $(\downprob, 0)$ and lie strictly between these lines
in $(0,\downprob)$. Suppose, to the contrary, that $v =\vtrans(u) \leq \downprob - u$ for $u \in
(0,\downprob)$. This would imply that the dual point $u^*$ given by
$u^* = u +v$ satisfies $u^* \leq \downprob$ which contradicts the
requirement that $u^*$ must be on the opposite side of the
fulcrum. Similarly,
if $v =\vtrans(u) \geq 1 - u$ for $u \in
(0,\downprob)$ then $u^* \geq 1$ and this is also not possible; if
$u^*=1$ then $u=0$ which is a contradiction.

Thus the curve that links $(0,1)$ and $(\downprob, 0)$ must take the
form
$$
\vtrans(u) = 
(\downprob -u) \Psi\left(\frac{u}{\downprob}\right) + (1-u) \left(1 -
  \Psi\left(\frac{u}{\downprob}\right) \right) =
(1-u) - (1-\downprob) \Psi \left( \frac{u}{\downprob} \right)
$$
where $\Psi(0) =0$, $\Psi(1)=1$ and $0 < \Psi(x) < 1$ for $x \in
(0,1)$. Clearly $\Psi$ must be continuous to satisfy the conditions of
the v-transform. It must also be strictly increasing. If it were not
then the derivative would satisfy $\vtrans^\prime(u) \geq -1$ which is
not possible: if at
any point $u \in
(0,\downprob)$ we have $\vtrans^\prime (u) = -1$ then the opposite
branch of the v-transform would have to jump vertically at the dual
point $u^*$, contradicting continuity; if $\vtrans^\prime (u) > -1$
then $\vtrans$ would have to be a decreasing function at $u^*$, which
is also a contradiction.

Thus $\Psi$ fulfills the conditions of a
continuous, strictly increasing distribution function on $[0,1]$ and
we have established the necessary form for the left branch
equation. To find the value of the right branch equation at $u >
\downprob$ we invoke the square property. Since $\vtrans(u) =
\vtrans(\udual) = \vtrans(u - \vtrans(u))$ we need to solve the
equation $x = \vtrans(u-x)$ for $x\in [0,1]$ using the formula for the
left branch equation of $\vtrans$. Thus we solve
$x = 1- u + x - (1-\downprob) \Psi(\tfrac{u-x}{\downprob})$ for $x$
and this yields the right branch equation as asserted.

\subsection{Proof of
  Proposition~\ref{prop:reconstructT}}\label{propr:proof-reconstructT}

Let $g_T(x)$ be as given in~\eqref{eq:volprofile} and let $u(x) = F_X(\mu_T-x)$. For $x \in \R^+$,  
$u(x)$ is a continuous, strictly decreasing function of $x$ starting
at $u(0) = \downprob$ and decreasing to $0$. Since $\Psi$ is a cumulative
distribution function, it follows that
\begin{displaymath}
  \udual(x) = u(x) + \vtrans \left(u(x)\right) = 1 - (1-\downprob)\Psi\left(\frac{u(x)}{\downprob}\right)
\end{displaymath}
is a continuous, strictly increasing function starting at $\udual(0) =
\downprob$ and increasing to $1$.
Hence $g_T(x) = F_X^{-1}(\udual(x)) - \mu_T$ is continuous
and strictly
increasing on $\R^+$ with $g_T(0) = 0$ as required of the profile
function of a volatility proxy transformation. It remains to check
that if we insert~\eqref{eq:volprofile} in~\eqref{eq:4} we recover $\vtrans(u)$, which is straightforward.


\subsection{Proof of Theorem~\ref{prop:copula-vtransform}}\label{prop:proof-copula-vtransform}
\begin{enumerate}
\item 
For any $0 \leq v \leq 1$ the event $\{U \leq u, V \leq v\}$ has zero
  probability for $u < \vtrans^{-1}(v)$. For $u \geq \vtrans^{-1}(v)$ we
  have
  \begin{displaymath}
    \{U \leq u, V \leq v\} = \{\vtrans^{-1}(v) \leq U \leq \min(u,\vtrans^{-1}(v)+v)\}
  \end{displaymath}
and hence $\P\left(U \leq u, V \leq v\right) =
\min(u,\vtrans^{-1}(v)+v) - \vtrans^{-1}(v) $ and~\eqref{eq:6} follows.
\item We can write $\P\left( U \leq u, V \leq v\right ) =
  C(u,v)$ where $C$ is the 
  copula given by~\eqref{eq:6}. It follows
  from the basic properties of a copula that
  \begin{equation*}
    \P\left( U \leq u, V = v\right ) = \frac{\rd}{\rd v}
    C(u,v) = 
\begin{cases}
0 & u < \vtrans^{-1}(v) \\
 - \frac{\rd}{\rd v} \vtrans^{-1}(v) & \vtrans^{-1}(v) \leq u < \vtrans^{-1}(v)+v \\
1 & u \geq \vtrans^{-1}(v)+v
\end{cases}
  \end{equation*}
This is the distribution function of a binomial distribution and it
must be the case that $ \Downprob(v)  =  - \frac{\rd}{\rd v}
\vtrans^{-1}(v)$. Equation~\eqref{eq:7} follows by differentiating the inverse.
\item Finally, $\E\left(\Downprob(V)\right) = \downprob$ is easily verified by
making the substitution $x =
\vtrans^{-1}(v)$ in the integral
                                             $ \E\left(\Downprob(V)\right)
                                              = - \int_0^1
                                              \frac{1}{\vtrans^\prime(\vtrans^{-1}(v))}
                                              \rd v $.
\end{enumerate}

\subsection{Proof of Proposition~\ref{prop:resconstructU}}\label{propr:proof-resconstructU}

It is obviously true that $\vtrans(\bm{\vtrans}^{-1}(v,W)) = v$ for any
$W$. Hence $\vtrans(U) = \vtrans(\bm{\vtrans}^{-1}(V,W)) = V$.
The uniformity of $U$ follows from the fact that
\begin{displaymath}
  \P\left(\bm{\vtrans}^{-1}(V,W) = \vtrans^{-1}(v) \mid V = v \right) =
  \P\left(W \leq \Delta(v) \mid V = v \right) = \P(W \leq \Delta(v) )= \Delta(v)\;.
\end{displaymath}
Hence the pair of random variables $(U,V)$ has the conditional distribution~(\ref{eq:35}) and is distributed according to
the copula $C$ in~(\ref{eq:6}).
%


\subsection{Proof of Theorem~\ref{theorem:multivariate-vtransform}}\label{theorem:proof-multivariate-vtransform}
\begin{enumerate}
\item
Since the event $\{V_i \leq v_i\}$ is equal to the event
$\{\vtrans^{-1}(v_i) \leq U_i \leq \vtrans^{-1}(v_i) + v_i \}$ we first compute the probability of a box $[a_1,b_1] \times \cdots \times [a_d, b_d]$
where $a_i = \vtrans^{-1}(v_i) \leq \vtrans^{-1}(v_i) + v_i =
b_i$. The standard formula for such probabilities implies that the
copulas
$C_{\bm{V}}$ and $C_{\bm{U}}$ are related by
 \begin{equation*}
 C_{\bm{V}}(v_1, \ldots, v_d) = \sum_{j_1=1}^2 \cdots \sum_{j_d=1}^2  (-1)^{j_1+ \cdots + j_d} C_{\bm{U}}(u_{1j_1}, \ldots,u_{d j_d})\;;
 \end{equation*}
see, for example,~\cite{bib:mcneil-frey-embrechts-15}, page 221. Thus
the copula densities are related by
 \begin{equation*}
 c_{\bm{V}}(v_1, \ldots, v_d) = \sum_{j_1=1}^2 \cdots \sum_{j_d=1}^2
c_{\bm{U}}(u_{1j_1}, \ldots,u_{d j_d})
 \prod_{i=1}^d \frac{\rd }{\rd v_i} (-1)^{j_i} u_{i j_i}
 \end{equation*}
and the result follows if we use~(\ref{eq:7}) to calculate that
 \begin{displaymath}
    \frac{\rd}{\rd v_i} (-1)^j u_{i j} =
\begin{cases}
  \frac{\rd}{\rd v_i}
    \left(- \vtrans^{-1}(v_i) \right)
    = \Downprob(v_i) & \text{if $j=1$,}\\
 \frac{\rd}{\rd v_i}
    \left(v_i + \vtrans^{-1}(v_i)\right) 
    = 1-\Downprob(v_i) & \text{if $j=2$.}
\end{cases}
 \end{displaymath}
\item
For the point $(u_1,\ldots,u_d) \in [0,1]^d$ we consider the set of events $A_i(u_i)$ defined by
\begin{equation*}
A_i(u_i) =
\begin{cases}
\left\{ U_i \leq u_i\right\}& \text{if $u_i \leq \downprob$} \\ 
\left\{ U_i > u_i\right\}& \text{if $u_i > \downprob$} 
\end{cases}
\end{equation*}
The probability $\P(A_1(u_1),\ldots,A_d(u_d))$ is the probability of
an orthant defined by the point $(u_1,\ldots,u_d)$ and the copula 
density at this point is given by
$$
c_{\bm{U}}(u_1,\ldots,u_d) = (-1)^{\sum_{i=1}^d \indicator{u_i > \downprob}}\frac{\rd^d}{\rd u_1 \cdots \rd u_d} \P\left(\bigcap_{i=1}^d A_i(u_i)\right)\;\;.
$$
The event $A_i(u_i)$ can be written
\begin{equation*}
A_i(u_i) =
\begin{cases}
\left\{ V_i \geq \vtrans(u_i), W_i \leq \Downprob(V_i) \right\}& \text{if $u_i \leq \downprob$} \\ 
\left\{ V_i > \vtrans(u_i), W_i > \Downprob(V_i) \right\}& \text{if $u_i > \downprob$} 
\end{cases}
\end{equation*}
and hence we can use Theorem~\ref{prop:copula-vtransform} to write
$$
\P\left( \bigcap_{i=1}^d A_i(u_i) \right) = \int_{\vtrans(u_1)}^1 \cdots \int_{\vtrans(u_d)}^1 c_{\bm{V}}(v_1,\ldots,v_d) \prod_{i=1}^d \Downprob(v_i)^{\indicator{u_i \leq \downprob}}(1-\Downprob(v_i))^{\indicator{u_i > \downprob}}
\rd v_1\cdots \rd v_d\;.
$$
The derivative is given by
$$
\frac{\rd^d}{\rd u_1 \cdots \rd u_d} \\P\left( \bigcap_{i=1}^d A_i(u_i) \right) = (-1)^d c_{\bm{V}}(\vtrans(u_1),\ldots,\vtrans(u_d)) \prod_{i=1}^d p(u_i)^{\indicator{u_i \leq \downprob}}(1-p(u_i))^{\indicator{u_i > \downprob}} \vtrans^\prime(u_i)
$$
where $p(u_i) = \Downprob(\vtrans(u_i))$ and hence we obtain
$$
c_{\bm{U}}(u_1,\ldots,u_d)  =  c_{\bm{V}}(\vtrans(u_1),\ldots,\vtrans(u_d)) \prod_{i=1}^d (-p(u_i))^{\indicator{u_i \leq \downprob}}(1-p(u_i))^{\indicator{u_i > \downprob}} \vtrans^\prime(u_i).
$$

It remains to verify that each of the terms in the product is identically equal to 1. 
For $u_i \leq \downprob$ this follows easily from~\eqref{eq:7} since $-p(u_i) = -\Downprob(\vtrans(u_i)) = 1/\vtrans^\prime(u_i)$. For $u_i >\downprob$ we need an expression for the derivative of the right branch equation. Since $\vtrans(u_i) = \vtrans(u_i - \vtrans(u_i))$ we obtain
\begin{displaymath}
\vtrans^\prime(u_i) = \vtrans^\prime(u_i - \vtrans(u_i))(1 - \vtrans^\prime(u_i)) = \vtrans^\prime(\udual_i)(1 - \vtrans^\prime(u_i))
\Longrightarrow \vtrans^\prime(u_i) = \frac{\vtrans^\prime(\udual_i)}{1+\vtrans^\prime(\udual_i)}
\end{displaymath}
implying that
\begin{displaymath}
1-p(u_i ) = 1-\Downprob(\vtrans(u_i)) = 1-\Downprob(\vtrans(\udual_i)) = 1 + \frac{1}{\vtrans^\prime(\udual_i)} = \frac{1+\vtrans^\prime(\udual_i)}{\vtrans^\prime(\udual_i)} = \frac{1}{\vtrans^\prime(u_i)}\;.
\end{displaymath}
\end{enumerate}

\subsection{Proof of Proposition~\ref{theorem:uncond-copula}}\label{theorem:proof-uncond-copula}
Let $V_t = \vtrans(U_t)$ and $Z_t = \Phi^{-1}(V_t)$ as usual. The
process $(Z_t)$ is an ARMA process with acf $\rho(k)$ and hence $(Z_{t_1},\ldots,Z_{t_k})$ are jointly
standard normally distributed with correlation matrix $P(t_1,\ldots,t_k)$. This implies
that the joint distribution function of $(V_{t_1},\ldots,V_{t_k})$ is the
Gaussian copula with density $c^{\text{Ga}}_{P(t_1,\ldots,t_k)}$ and hence by Part 2 of Theorem~\ref{theorem:multivariate-vtransform} 
the joint distribution function of $(U_{t_1},\ldots,U_{t_k})$ is the copula with density $c^{\text{Ga}}_{P(t_1,\ldots,t_k)}(\vtrans(u_{1}),\ldots,\vtrans(u_{k}))$.

\subsection{Proof of Proposition~\ref{prop:ARMA-dependence}}\label{prop:proof-ARMA-dependence}
We split the integral in~(\ref{eq:29}) into four parts. First observe
that by making the substitutions $v_1 = \vtrans(u_1) = 1-u_1/\downprob$ and $v_2
= \vtrans(u_2) = 1-u_2/\downprob$ on $[0,\downprob] \times [0,\downprob]$ we get
\begin{align*}
\int_0^\downprob \int_0^\downprob u_1 u_2 c^{\text{Ga}}_{\rho(k)} \left(\vtrans(u_1),
  \vtrans(u_2)\right) \rd u_1 \rd u_2
  &= \downprob^4\int_0^1 \int_0^1 (1-v_1)(1-v_2)
                                       c^{\text{Ga}}_{\rho(k)}\left(v_1,v_2
                                       \right) \rd v_1 \rd v_2 \\
&= \downprob^4 \E( (1-V_{t})(1-V_{t+k}) ) \\
&= \downprob^4 \left( 1 - \E(V_t) - \E(V_{t+k}) + \E(V_{t}V_{t+k}) \right) = \downprob^4 \E(V_{t}V_{t+k})
\end{align*}
where $(V_t,V_{t+k})$ has joint distribution given by
the Gaussian copula $C^{\text{Ga}}_{\rho(k)}$. 
Similarly
by making the substitutions $v_1 = \vtrans(u_1) = 1-u_1/\downprob$ and $v_2
=\vtrans(u_2)=(u_2-\downprob)/(1-\downprob)$ on $[0,\downprob] \times [\downprob,1]$ we get
 \begin{align*}
 \lefteqn{\int_0^\downprob \int_\downprob^1 u_1 u_2 c^{\text{Ga}}_{\rho(k)}\left(\vtrans(u_1),
   \vtrans(u_2)\right) \rd u_1 \rd u_2} \\  &= \int_0^1 \int_0^1
                                            \downprob^2(1-\downprob)(1-v_1)
 \Big(\downprob + (1-\downprob)v_2 \Big)
                                       c^{\text{Ga}}_{\rho(k)}\left(v_1,v_2
                                        \right) \rd v_1 \rd v_2 \\
 &= \downprob^3 (1-\downprob)\E(1-V_t) + \downprob^2
   (1-\downprob)^2\E\left((1-V_t)V_{t+k}\right)= \frac{\downprob^2(1-\downprob)}{2} - \downprob^2(1-\downprob)^2 \E(V_t V_{t+k})
 \end{align*}
 and the same value is obtained on the quadrant $[\downprob,1] \times
 [0,\downprob]$. 
Finally making the substitutions $v_1 = \vtrans(u_1) = (u_1-\downprob)/(1-\downprob)$ and $v_2
=\vtrans(u_2)=(u_2-\downprob)/(1-\downprob)$ on $[\downprob,1] \times [\downprob,1]$ we
get
 \begin{align*}
\lefteqn{\int_\downprob^1 \int_\downprob^1 u_1 u_2 c^{\text{Ga}}_{\rho(k)}\left(\vtrans(u_1),
  \vtrans(u_2)\right) \rd u_1 \rd u_2}\\
 &= \int_0^1 \int_0^1    (1-\downprob)^2
 \Big(\downprob + (1-\downprob)v_1 \Big) \Big(\downprob + (1-\downprob)v_2 \Big)
                                       c^{\text{Ga}}_{\rho(k)}\left(v_1,v_2
                                        \right) \rd v_1 \rd v_2
 \\
&= \int_0^1 \int_0^1
                                           (1-\downprob)^2
\Big( \downprob^2 + \downprob(1-\downprob) v_1 + \downprob(1-\downprob) v_2 + (1-\downprob)^2
  v_1 v_2\Big)
                                       c^{\text{Ga}}_{\rho(k)}\left(v_1,v_2
      \right) \rd v_1 \rd v_2
 \end{align*}
 \begin{align*}
&=\downprob^2(1-\downprob)^2 + \downprob (1-\downprob)^3 \E(V_t) + \downprob(1-\downprob)^3
  \E(V_{t+k}) + (1-\downprob)^4 \E(V_t V_{t+k}) \\
&= \downprob(1-\downprob)^2 + (1-\downprob)^4 \E(V_t V_{t+k})
 \end{align*}
Collecting all of these terms together yields
\begin{align*}
\int_0^1 \int_0^1 u_1 u_2 c^{\text{Ga}}_{\rho(k)}\left(\vtrans(u_1),
  \vtrans(u_2)\right) \rd u_1 \rd u_2
&= \downprob(1-\downprob) + (2\downprob-1)^2  \E(V_t
  V_{t+k})
\end{align*}
 and since $\rho_S(Z_t ,Z_{t+k}) = 12 \E(V_t V_{t+k}) -3$ it follows that
\begin{align*}
\rho(U_t, U_{t+k}) =12\E(U_t U_{t+k}) -3&= 12 \int_0^1 \int_0^1 u_1 u_2 c^{\text{Ga}}_{\rho(k)}\left(\vtrans(u_1),
  \vtrans(u_2)\right) \rd u_1 \rd u_2 -3 \\
&= 12 \downprob(1-\downprob) + 12 (2\downprob-1)^2  \E(V_t
  V_{t+k})  -3 \\ &=
12 \downprob(1-\downprob) +  (2\downprob-1)^2 \left( \rho_S(Z_t ,Z_{t+k}) +3
                    \right) -3 \\
&=  (2\downprob-1)^2 \rho_S(Z_t ,Z_{t+k})\,.
\end{align*}
The value of Spearman's rho $ \rho_S(Z_t ,Z_{t+k})$ for the bivariate
Gaussian distribution is well known; see for
example~\cite{bib:mcneil-frey-embrechts-15}.


\subsection{Proof of
  Proposition~\ref{theorem:cond-density}}\label{theorem:proof-cond-density}
The conditional density satisfies
\begin{eqnarray*}
  f_{U_t \mid \bm{U}_{t-1}}(u \mid \bm{u}_{t-1})  & = &
                                                        \frac{c_{\bm{U}_t}(u_1,\ldots,u_{t-1},u)}{c_{\bm{U}_{t-1}}(u_1,\ldots,u_{t-1})}
                                                        =
                                                        \frac{ c^{\text{Ga}}_{P(1,\ldots,t)}(\vtrans(u_1),\ldots,\vtrans(u_{t-1}),\vtrans(u))}{c^{\text{Ga}}_{P(1,\ldots,t-1)}(\vtrans(u_1),\ldots,\vtrans(u_{t-1}))}\;.
\end{eqnarray*}
The Gaussian copula density is given in general by
\begin{displaymath}
  c^{\text{Ga}}_P(v_1,\ldots,v_d) =
  \frac{f_{\bm{Z}}\big(\Phi^{-1}(v_1),\ldots,\Phi^{-1}(v_d)\big)}{\prod_{i=1}^d \phi\big(\Phi^{-1}(v_i)\big)}
\end{displaymath}
where $\bm{Z}$ is a multivariate Gaussian vector with standard normal
margins and correlation matrix $P$. Hence it follows that we can write
\begin{eqnarray*}
  f_{U_t \mid \bm{U}_{t-1}}(u \mid \bm{u}_{t-1}) & = &
\frac{f_{\bm{Z}_t}\Big(  \Phi^{-1}\big(\vtrans(u_1)\big), \ldots,
                                                       \Phi^{-1}\big(\vtrans(u_{t-1})\big), \Phi^{-1}\big(\vtrans(u)\big)
                                                       \Big)}{
                                                       f_{\bm{Z}_{t-1}}\Big(  \Phi^{-1}\big(\vtrans(u_1)\big), \ldots,
                                                       \Phi^{-1}\big(\vtrans(u_{t-1})\big)
                                                       \Big)
                                                       \phi\big(
                                                       \Phi^{-1}\big(\vtrans(u)\big)
                                                       \big)}  \\                                                    
  & = & \frac{f_{Z_t
      \mid \bm{Z}_{t-1}}\Big( \Phi^{-1}\big(\vtrans(u)\big) \mid \Phi^{-1}\big(\vtrans(\bm{u}_{t-1})\big) \Big)}{  \phi\big(\Phi^{-1}\big(\vtrans(u)\big)\big) }
\end{eqnarray*}
where $f_{Z_t
      \mid \bm{Z}_{t-1}}$ is the conditional density of the ARMA
      process, from which~\eqref{eq:22} follows easily.

      \bibliographystyle{jf}

      \newcommand{\noopsort}[1]{}

\end{document}